\documentclass[%
 reprint,
superscriptaddress,
 amsmath,amssymb,
 aps,
]{revtex4-2}

\usepackage[utf8]{inputenc}
\usepackage{amsmath}
\usepackage{bm}
\usepackage{graphicx}
\usepackage{amssymb}
\usepackage{times}
\usepackage{color}
\usepackage{float}
\usepackage{multirow}
\usepackage{subcaption}
\usepackage{braket}
\usepackage{hyperref}
\usepackage{chemformula}
\usepackage{makecell}
\usepackage{xfrac} 
\usepackage{threeparttable}
\usepackage{booktabs}
\usepackage{comment}

\begin{document}

\title{Color Centers in Cubic Boron Nitride}

\author{William Stenlund}
\email{william.stenlund@liu.se}
\affiliation{Department of Physics, Chemistry and Biology, Link\"oping
 University, Link\"oping, Sweden}

\author{Joel Davidsson}
\affiliation{Department of Physics, Chemistry and Biology, Link\"oping University, Link\"oping, Sweden}

\author{Viktor Ivády}
\affiliation{MTA-ELTE Lendület “Momentum” NewQubit Research Group, 
Pázmány Péter, Sétány 1/A, 1117, Budapest, Hungary}
\affiliation{Department of Physics of Complex Systems, Eötvös Loránd University, Egyetem tér 1-3, H-1053 Budapest, Hungary}

\author{Rickard Armiento}
\affiliation{Department of Physics, Chemistry and Biology, Link\"oping University, Link\"oping, Sweden}

\author{Igor A. Abrikosov}
\affiliation{Department of Physics, Chemistry and Biology, Link\"oping University, Link\"oping, Sweden}

\begin{abstract}

\begin{comment}
Cubic boron nitride (c-BN) is a wide-bandgap semiconductor that hosts point defects with optical transitions in the telecom-wavelength, potentially leading to applications in quantum technologies such as quantum sensing or single photon emitters.
Multiple defect complexes have been found, however, they include impurity elements from a small section of the periodic table.
We apply density functional theory and a high-throughput methodology, ADAQ, to broadly screen for point defect complexes containing s- or p-elements. 
The X simulated defects were evaluated in terms of formation energy, ZPL and transition dipole moment for different charge and spin states.
\end{comment}

Cubic boron nitride (c-BN) is a wide-bandgap semiconductor (WBGS) with potential applications in both power electronics and quantum technologies. Color centers in WBGS can be used as single photon emitters and quantum sensors. Several zero phonon lines have been measured in c-BN experiment but not yet identified. 
To systematically probe the combinatorially complex chemical space of defects, we generate a large-scale point defect data set for c-BN. We apply density functional theory calculation implemented in a high-throughput workflow Automatic Defect Analysis and Qualification (ADAQ) to broadly screen for point defect complexes containing s- or p-elements. More than 8000 defects have been calculated in different charge and spin states. The calculated properties are stored in defect database and are then filtered to find defects with properties similar to the NV-center in diamond. More accurate calculations using hybrid functionals are then performed on a selected set of promising defects to further assess their suitability for quantum technology. In particular, we reexamined the $\mathrm{O_NV_B}$ defect which likely explains the GC-2 line. The hybrid calculations also suggest other defect candidates with bright emission, such as two carbon defects and the $\mathrm{Na_B^-}$ defect.

\end{abstract}

\maketitle

%%%%%%%%%%%%%%%%%%%%%%%%%%%%%%%%%%%%%%%%%%%%%%%%%%%%%%%%%%%%%%%%%%%%%%%%%%%%%%%%%%%%%%%%%%%%%%%%%%%%%%%%
%%%%%%%%%%%%%%%%%%%%%%%%%%%%%%%%%%%%%%%%%%%%%%%%%%%%%%%%%%%%%%%%%%%%%%%%%%%%%%%%%%%%%%%%%%%%%%%%%%%%%%%%

\section{Introduction}

Defects in wide-bandgap semiconductors (WBGS) can serve as a platform for quantum technologies, exemplified by the $\mathrm{NV^-}$ center in diamond which has found use as a defect qubit \cite{Nizovtsev2005, doi:10.1073/pnas.1003052107} and as quantum sensors for temperature \cite{doi:10.1021/nl401216y}, magnetic field \cite{Hong_2013} and strain \cite{doi:10.1021/acs.nanolett.9b01402} at room temperature.
Diamond has many useful material properties and applications that are thoroughly studied. Cubic boron nitride (c-BN) is isoelectronic with diamond, thus, similar properties are found. Hexagonal boron nitride (h-BN) is a layered material like graphite, it is also a WBGS which hosts several emitters that have been investigated both theoretically \cite{PhysRevB.106.014107} and experimentally \cite{doi:10.1021/acsnano.4c03640}. In particular, the negatively charged boron vacancy center ($\mathrm{V_B^-}$) defect \cite{Ivady2020, Gottscholl2020} is showing promise as a sensor for temperature \cite{doi:10.1021/acsphotonics.1c00320, Gottscholl2021}, magnetic fields \cite{doi:10.1021/acs.nanolett.3c00849, Gottscholl2021} and strain \cite{doi:10.1021/acs.nanolett.2c01722, Gottscholl2021}, similar to the NV-center in diamond.

Despite the great interest in h-BN, c-BN remains comparatively underexplored as a host for quantum defects. There has been no systematic study of point defects in c-BN. Previously simulated defects are intrinsic defects \cite{Nguyen2025} or impurities with oxygen, carbon or silicon, presented in Table~\ref{tab:knowntheory}. These defects are simulated with density functional theory (DFT) using the hybrid functional by Heyd, Scuseria, and Ernzerhof (HSE06) \cite{HSE, HSEerrata} to better model the band gap of c-BN and the zero phonon line (ZPL) of the defects. Note that different values of the exact exchange mixing parameters $\alpha$ can be used, 25\% is the default $\alpha$ value for the HSE functional \cite{HSE, HSEerrata}, while 33\% can be used to better match the calculated band gap with the experimental gap of c-BN.  
The previous theoretical works presented in Table~\ref{tab:knowntheory} examine defects consisting of a narrow (albeit commonly occurring) set of elements, this work intends to screen a larger set of impurity elements.

The experimental indirect band gap of c-BN is 6.4 eV \cite{CHRENKO1974511} and there are several emission lines found in experiment \cite{PhysRevB.98.094106, Lopez-Morales:20}. However, most of these emission lines do not have a specific attribution. The neutral $\mathrm{O_NV_B}$ defect is known theoretically \cite{PhysRevLett.113.136401, NGUYEN2025113388} and is proposed \cite{PhysRevLett.113.136401} to be the cause of the GC-2 emission line at 1.63 eV \cite{Tkachev1985, Shipilo1986}.  Zero field splitting and hyperfine splitting parameters have also been calculated for the $\mathrm{O_NV_B}$ defect \cite{10.1063/1.5083076}.

In this work, we explore possible defects in c-BN with the goal of finding bright emitters and assessing their suitability for different quantum applications. In particular, defects with spin that emit brightly in the telecom (1260-1675 nm) and biological (650-1350 nm) wavelength ranges are of interest. Defects emitting light at around 500-650 nm where there are known experimental lines \cite{PhysRevB.98.094106, Lopez-Morales:20} are also of interest. 

Section \ref{sec:high-throughput} presents the methodology and results of the high-throughput screening of defects in c-BN. A subset of defects are simulated with HSE06 and section \ref{sec:res} presents and discusses these results. Lastly, the conclusions of this study are summarized in \ref{sec:conc}.

\begin{table}[h!]
\caption{Previously theoretically investigated defects in c-BN and their corresponding calculated ZPL. The hybrid functional used the percentage of exact exchange mixing parameters $\alpha$ are also noted.}
\begin{tabular} {c|c|c|c|l}
Defect & Charge & Spin & ZPL [eV] & Functional, $\alpha$ \\

\hline
 $\mathrm{O_NV_B}$ & 0 & 1 & 1.60 & HSE, unspecified \cite{PhysRevLett.113.136401}\\
 $\mathrm{C_BV_B}$  &  0  &  1  &  0.95 & HSE, 33\% \cite{PhysRevB.108.L041102} \\
 $\mathrm{Si_BV_B}$ &  0  &  1  &  0.89 &  HSE, 33\% \cite{PhysRevB.108.L041102} \\
 %$\mathrm{V_BC_B}$ & 0 & 1 &  0.96 &  DDH, 27\% \cite{NGUYEN2025113388}\\
 %$\mathrm{V_BSi_B}$ & 0 & 1 &  0.96 &  DDH, 27\% \cite{NGUYEN2025113388}\\
 %$\mathrm{V_BGe_B}$ & 0 & 1 &  0.97 &  DDH, 27\% \cite{NGUYEN2025113388}\\
 %Dielectric dependent hybrid (DDH) \cite{PhysRevB.93.235106} functionals sets EXX to $\alpha = \frac{1}{\epsilon_{\infty}}$.
\hline
 
\end{tabular}
\label{tab:knowntheory}
\end{table}

\section{High-Throughput Screening}
\label{sec:high-throughput}
\subsection{High-Throughput Calculations}

The workflows ADAQ \cite{ADAQ}, which uses the high-throughput toolkit httk \cite{armientoDatabaseDrivenHighThroughputCalculations2020}, was used to generate 8625 defects, including substitutional defect, vacancies, interstitials, and their complexes, in c-BN. 
These defects are placed in a cubic 512 atom (4x4x4) supercell.
Extrinsic dopants are limited to the s- and p-elements, up to bismuth excluding noble gases except for xenon which is a known emitter in diamond \cite{MARTINOVICH2003785, SANDSTROM2018182}. Only single and double defects are considered, but not interstitial-interstitial pairs. The generated defects were simulated with a lower accuracy screening workflow in the ADAQ framework which uses DFT, see further details in Section~\ref{sec:DFTPBE}. The defects are simulated at different charge and spin states, including different excitations. If the eigenvalue difference between the lowest unoccupied and the highest occupied orbitals is too low, in this case below 0.4 eV, the second lowest excitation is considered instead. Based on the DFT simulations the following properties are calculated: formation energy, zero-phonon line (ZPL) and transition dipole moment (TDM), which are described in more detail in Ref.~\onlinecite{ADAQ}. Beyond these, $\Delta Q$ is a measure of the geometry difference between the ground and excited states, weighted by atomic mass, it is calculated as follows:

\begin{equation}
\label{eq:dq}
    (\Delta Q )^2= \sum_a m_a |\vec{R_{es,a}}-\vec{R_{gs,a}}|^2 , 
\end{equation} where $m_a$ is the mass of atom $a$, and $\vec{R_{gs,a}}$ and $\vec{R_{es,a}}$ are the positions of atom $a$ in the ground state (gs) and the excited state (es) respectively.

The Huang-Rhys and Debye-Waller factors are calculated using $\Delta Q$ in the one-phonon approximation \cite{Alkauskas_2014}. These calculated properties are stored in the ADAQ database \cite{adaq_database}.

\begin{figure}[H]
  \includegraphics[width=\columnwidth]{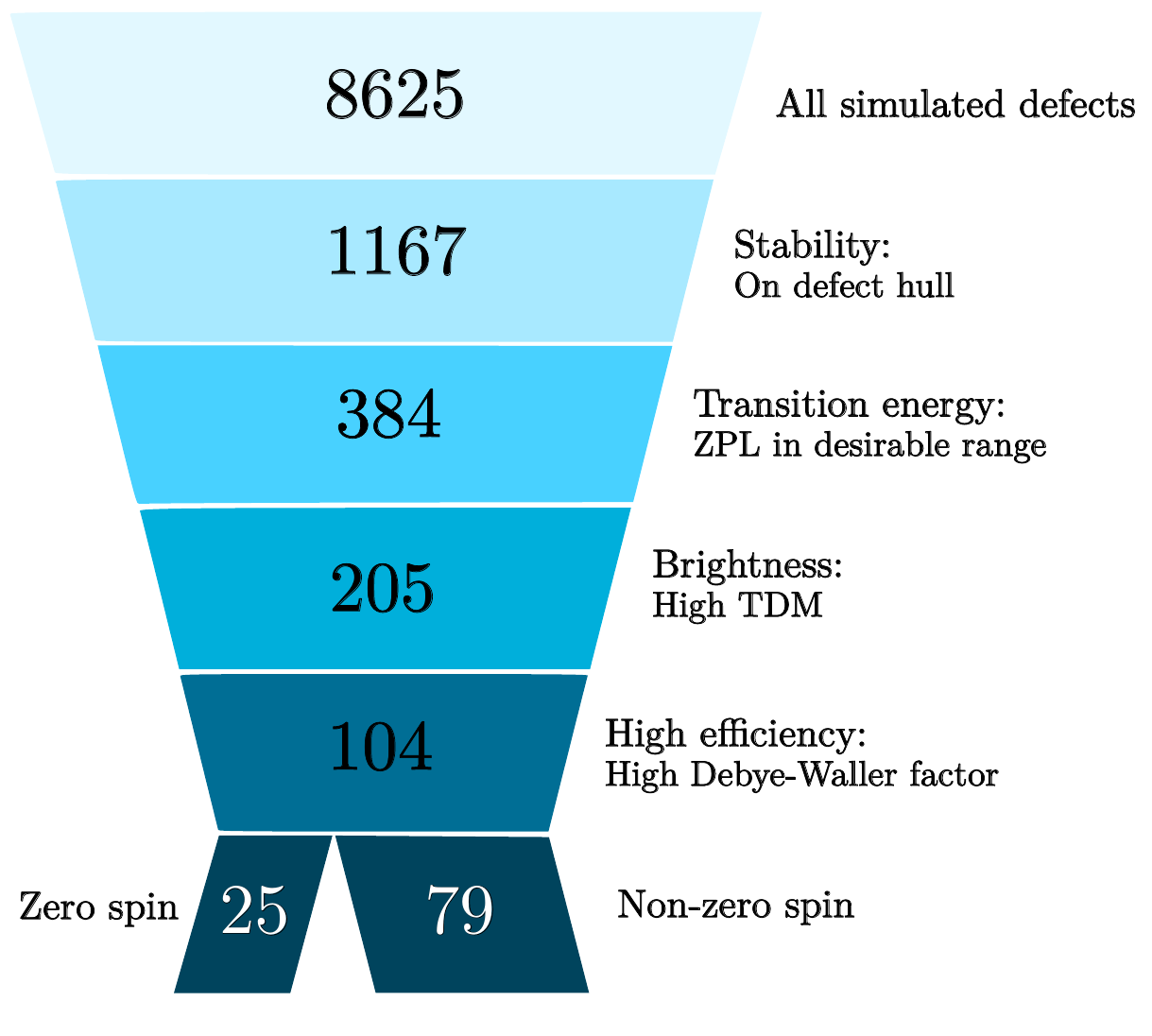}
	\caption{Schematic description of the screening process. All simulated defects are present at the top and are then gradually filtered away when failing certain criteria. At the bottom, 79 defects with non-zero spin and 25 defects with spin-0 are shown. They have passed the screening.}
	\label{fig:screen} 
\end{figure}

\subsection{Methodological Details of Low Accuracy Screening Workflow}
\label{sec:DFTPBE}

The generated defects were simulated with DFT and the calculations are performed with VASP 5.4.4 \cite{VASP,VASP2}, which uses the projector-augmented wave method \cite{PAW, Kresse99}. The semi-local exchange correlation functional by Perdew, Burke, and Ernzerhof (PBE) \cite{PBE} was used for these high-throughput calculations. The defects were simulated at the $\Gamma$-point, convergence tests indicated that the structures are sufficiently converged for the $\Gamma$-point with a 512 atom supercell. The lattice parameter used is 3.6246 Å, which is the equilibrium lattice parameter for pure c-BN calculated with the PBE functional. Excited states are simulated with the constrained occupation method \cite{PhysRevLett.103.186404}. The geometry of the supercells with the defects are relaxed without symmetry constraints in both ground and excited states. The plane wave cutoff energy is 600 eV. The stopping parameters are $1\cdot10^{-4}$ eV for electronic relaxation and $5 \cdot 10^{-3}$ eV for ionic relaxation.
Formation energies are calculated with the Lany-Zunger charge correction \cite{PhysRevB.78.235104}.

\subsection{Database Screening}
\label{sec:databasescreen}
The generated database of defects in c-BN is searched for defects with suitable properties for single photon emission or solid state qubit applications, according to the criteria displayed in Figure~\ref{fig:screen}. The first step is to screen for defects that are thermodynamically stable, that is, defects that have a formation energy that lies on the defect hull \cite{DavidssonBabarShafizadehIvanovIvadyArmientoAbrikosov+2022+4565+4580} for its particular defect stoichiometry.

\begin{figure}[]
  \includegraphics[width=\columnwidth]{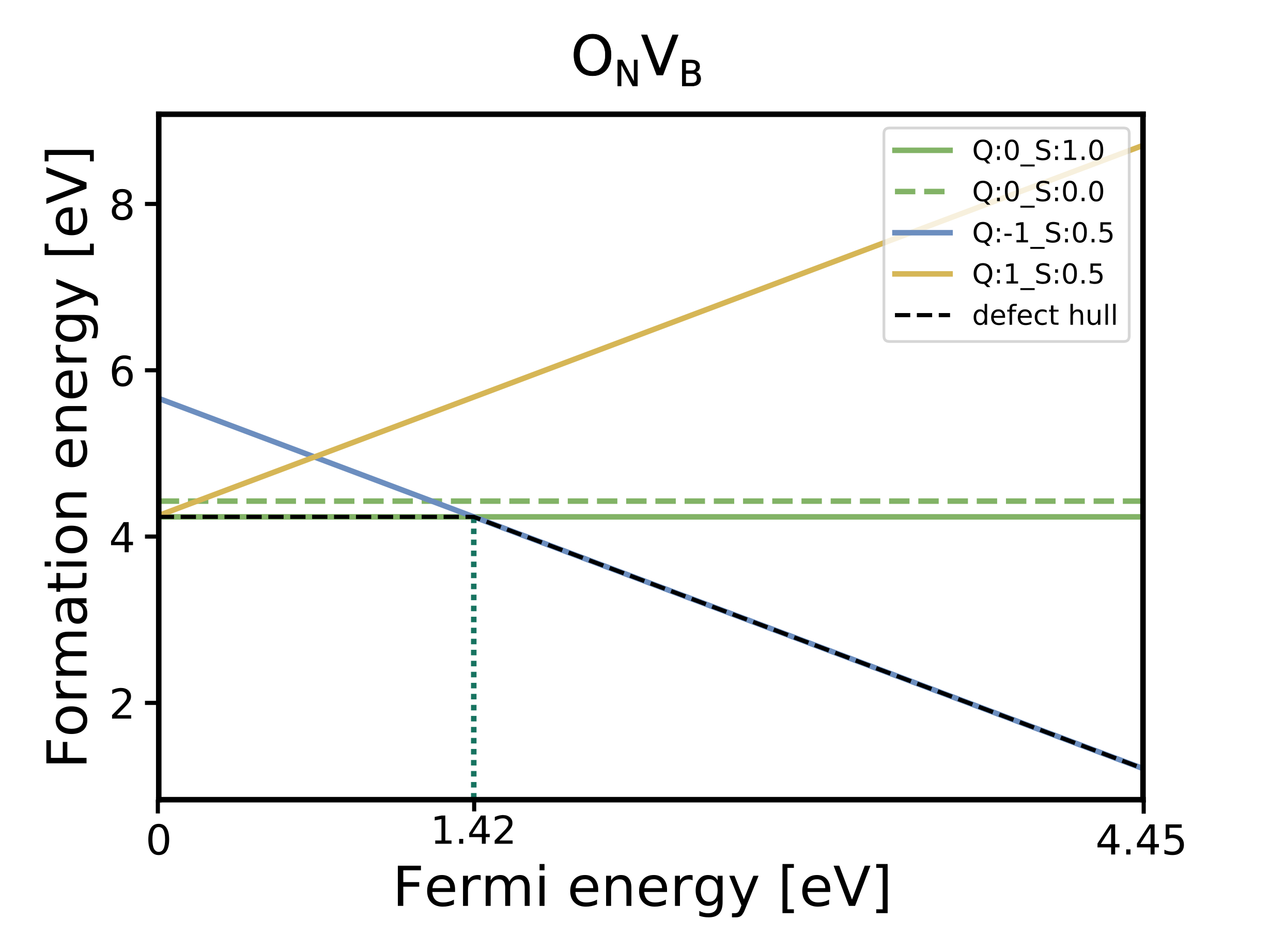}
	\caption{Formation energy diagram of the different charge (Q) and spin (S) configurations of the $\mathrm{O_NV_B}$ defect, calculated from the ADAQ screening workflow. The neutral and negative charge states are on the convex hull and the Fermi energy where the neutral and positive charge have the same formation energy, called the charge transition level $\epsilon(+/0)$, is 1.42 eV. When using the PBE functional, the simulated band gap is 4.45 eV.}
	\label{fig:onvb_form} 
\end{figure}

For example, defects with an extra oxygen atom, a missing N atom and a missing B atom in the supercell can exist in multiple configurations, spin and charge states. The formation energy will vary with the Fermi energy, except for neutral defects. The configurations that have the lowest formation energy for a range of the Fermi energies are said to be on the defect hull. Figure~\ref{fig:onvb_form} shows an example of the $\mathrm{O_NV_B}$ defect where the formation energy diagram indicates that only the spin-1 neutral state and the spin-1/2 negative state are on the defect hull, the spin-0 neutral is not on the hull as it has higher formation energy than the spin-1 neutral state. The figure also shows that the charge transition fermi energy $\epsilon(+/0)$ is 1.42 eV above the VBM, this is likely an underestimation due to the ADAQ simulations using the PBE functional.

The defects that passed the first step are then assessed by their optical properties. For different applications different transition energies may be desirable. To filter out defects that emit at lower energy, we require the ZPL calculated with the PBE functional to be greater than 0.5 eV, as non-radiative recombination is expected to be the dominant mechanism for lower energies. To find defects whose emission is likely to be observable, we search for a high TDM, resulting in a high intensity, and a high Debye-Waller factor so a significant amount of light is emitted in the ZPL rather than the phonon sideband. In practice, $\Delta$Q is used for screening instead of the Debye-Waller factor, since $\Delta$Q is the dominant factor as it scales to the power of 2.
Finally, the spin of the defects is considered since it affects the interaction with magnetic fields. Non-zero spin can result in a two-level system suitable for qubits. Defects that pass this screening are summarized in Table~\ref{tab:screened}. Some of these defects are selected for further investigation and are simulated with the HSE functional to get a more accurate estimation of their properties.

\subsection{Results of Database Filtering}
\label{sec:database}

In total, 8625 defects in c-BN were simulated with the ADAQ screening workflow. 1167 of these defects have a ground state on the defect hull. Of these defects 118 did not have any local excitations. For the remaining defects, calculations of excited states have converged for 492 defects, or for 47\% of the considered cases. 

Next, we consider the optical properties; 384 defects had a ZPL above 0.5 eV, 205 had a TDM above 3 Debye and 104 had a $\Delta$Q $<$ 1 amu$^{1/2}$Å. Lastly, spin is considered, 79 defects also have non-zero spin, and these are presented in Table~\ref{tab:screened}.

\begin{table*}[htb]
\caption{Summary of the defects in c-BN simulated with the ADAQ screening workflow. 79 stable defects with non-zero spin and good optical properties are shown. Vacancies and substitutional atoms $X$ on boron and nitrogen sites are denoted by $\mathrm{V_B}$, $\mathrm{V_N}$, $\mathrm{X_B}$ and $\mathrm{X_N}$. An interstitial X atom is denoted by $\mathrm{Int_X}$. Defects investigated further with the HSE functional are marked with bold text.}
\setlength{\tabcolsep}{4pt}
\begin{minipage}{\columnwidth}
\centering
\begin{tabular} {c|cc|ccc}
Defect & Charge & Spin & ZPL & TDM & $\Delta$Q \\
 &  &  &  eV & Debye & amu$^{1/2}$Å \\
\hline
$\mathrm{Cl_BInt_{Cl}}$  &   0  &  0.5  &  2.36  &  6.27  &  0.7  \\
$\mathrm{Cl_BCl_B}$  &    -1  &  0.5  &  0.81  &  3.41  &  0.32 \\
$\mathrm{Cl_BCl_B}$    &  1  &  1.5  &  0.76  &  11.82  &  0.47  \\
$\mathrm{B_NK_N}$   &  -1  &  0.5  &  0.67  &  10.62  &  0.24  \\
$\mathrm{B_NK_N}$    &  1  &  0.5  &  0.68  &  8.14  &  0.41\\
$\mathrm{Int_{Be}N_B}$   &  1  &  0.5  &  1.21  &  4.01  &  0.43  \\
$\mathbf{O_NV_B}$   &  0  &  1.0  &  1.1  &  5.07  &  0.39 \\
$\mathbf{O_NV_B}$    &  -1  &  0.5  &  1.13  &  9.43  &  0.46  \\
$\mathrm{B_NBa_N}$    &  -1  &  1.0  &  0.54  &  7.5  &  0.32  \\
$\mathrm{F_BF_B}$    &  0  &  2.0  &  0.68  &  6.89  &  0.46  \\
$\mathrm{F_BF_B}$    &  1  &  1.5  &  0.65  &  7.48  &  0.23  \\
$\mathrm{Int_HV_B}$    &  -1  &  0.5  &  0.98  &  7.51  &  0.41  \\
$\mathrm{As_BV_B}$    &  1  &  1.0  &  0.61  &  8.97  &  0.42 \\
$\mathrm{Bi_BBi_B}$    &  -1  &  0.5  &  0.63  &  12.95  &  0.65 \\
$\mathrm{Int_FV_B}$    &  0  &  1.0  &  0.89  &  9.09  &  0.71 \\
$\mathrm{Sb_B}$    &  1  &  0.5  &  0.68  &  6.01  &  0.78 \\
$\mathrm{Ca_BInt_B}$    &  1  &  0.5  &  2.01  &  3.95  &  0.67 \\
$\mathrm{Ca_NV_N}$    &  1  &  0.5  &  0.94  &  6.9  &  0.95 \\
$\mathrm{Int_{Br}V_B}$   &  -1  &  0.5  &  0.67  &  9.03  &  0.75 \\
$\mathrm{Bi_BV_B}$  &  1  &  1.0  &  0.52  &  6.13  &  0.52 \\
$\mathrm{C_NV_N}$    &  1  &  0.5  &  0.66  &  14.74  &  0.95 \\
$\mathrm{B_NTl_B}$    &  -1  &  0.5  &  1.53  &  3.23  &  0.78 \\
$\mathrm{Int_NSr_B}$   &  -1  &  0.5  &  0.51  &  3.16  &  0.85 \\
$\mathrm{Int_{Be}N_B}$    &  1  &  0.5  &  1.21  &  4.01  &  0.43 \\
$\mathrm{Ba_NInt_B}$    &  1  &  0.5  &  0.57  &  3.87  &  0.59 \\
$\mathrm{Int_BTe_N}$    &  1  &  0.5  &  2.26  &  5.75  &  0.82 \\
$\mathrm{B_NV_B}$    &  0  &  0.5  &  3.69  &  6.26  &  0.91 \\
$\mathrm{Int_{Rb}Rb_N}$    &  -1  &  2.0  &  1.09  &  6.12  &  0.33 \\
$\mathrm{B_NCa_N}$    &  -1  &  1.0  &  0.59  &  6.53  &  0.33 \\
$\mathrm{Int_{Ba}}$    &  1  &  0.5  &  0.89  &  3.45  &  0.93 \\
$\mathrm{Int_SN_B}$    &  1  &  0.5  &  1.0  &  4.41  &  0.62 \\
$\mathrm{F_BV_B}$   &  0  &  1.5  &  0.52  &  9.34  &  0.4 \\
$\mathrm{F_BV_B}$    &  -1  &  2.0  &  0.53  &  5.54  &  0.69 \\
$\mathrm{B_NRb_N}$    &  -1  &  0.5  &  0.65  &  10.64  &  0.3 \\
$\mathrm{B_NRb_N}$    &  1  &  0.5  &  0.69  &  8.95  &  0.47 \\
$\mathbf{Na_B}$    &  -1  &  0.5  &  0.99  &  7.83  &  0.14 \\
$\mathrm{F_BInt_F}$    &  0  &  0.5  &  0.92  &  6.36  &  0.25 \\
$\mathrm{V_BV_N}$    &  1  &  1.5  &  2.73  &  6.78  &  0.91 \\
$\mathrm{Be_BV_N}$    &  1  &  0.5  &  3.0  &  6.39  &  0.82 \\
$\mathrm{Se_BSe_B}$    &  -1  &  0.5  &  1.38  &  4.29  &  0.91 \\
\end{tabular}
\end{minipage}
\begin{minipage}{\columnwidth}
\begin{tabular} {c|cc|ccc}
Defect & Charge & Spin & ZPL & TDM & $\Delta$Q \\
 &  &  &  eV & Debye & amu$^{1/2}$Å \\
\hline
$\mathrm{P_BV_B}$    &  1  &  1.0  &  0.74  &  8.0  &  0.43 \\
$\mathrm{Int_BRb_N}$    &  0  &  0.5  &  0.86  &  5.39  &  0.86 \\
$\mathrm{Sb_BV_B}$   &  1  &  1.0  &  0.54  &  8.69  &  0.44 \\
$\mathrm{Cl_BInt_{Cl}}$    &  0  &  0.5  &  2.36  &  6.27  &  0.7 \\
$\mathrm{S_BS_B}$    &  1  &  0.5  &  0.5  &  19.6  &  0.45 \\
$\mathrm{Int_SN_B}$    &  1  &  0.5  &  1.0  &  4.41  &  0.62 \\
$\mathrm{K_BK_N}$    &  1  &  0.5  &  0.72  &  7.9  &  0.47 \\
$\mathrm{B_NXe_N}$    &  0  &  0.5  &  0.58  &  8.22  &  0.36 \\
$\mathrm{B_NInt_N}$    &  0  &  0.5  &  2.25  &  5.73  &  0.8 \\
$\mathrm{Na_BV_B}$    &  -1  &  1.0  &  1.13  &  3.29  &  0.61 \\
$\mathrm{Int_NLi_B}$    &  0  &  0.5  &  0.61  &  11.17  &  0.35 \\
$\mathrm{B_NInt_{Se}}$    &  -1  &  0.5  &  0.76  &  3.11  &  0.8 \\
$\mathrm{B_NSr_N}$    &  -1  &  1.0  &  0.6  &  7.88  &  0.3 \\
$\mathrm{Ba_N}$    &  0  &  1.5  &  0.53  &  9.85  &  0.32 \\
$\mathrm{H_BN_B}$    &  -1  &  0.5  &  1.62  &  4.44  &  0.96 \\
$\mathrm{O_BV_B}$    &  1  &  0.5  &  0.55  &  8.52  &  0.39 \\
$\mathrm{Be_BN_B}$    &  -1  &  1.0  &  2.64  &  7.11  &  0.93 \\
$\mathbf{C_NV_B}$    &  -1  &  0.5  &  0.89  &  3.5  &  0.43 \\
$\mathrm{N_BV_B}$    &  0  &  0.5  &  0.98  &  7.36  &  0.43 \\
$\mathrm{N_BV_B}$    &  1  &  1.0  &  0.8  &  6.15  &  0.46 \\
$\mathrm{Sr_BV_B}$  &  0  &  1.0  &  0.51  &  4.92  &  0.38 \\
$\mathrm{N_BNa_B}$    &  1  &  0.5  &  0.62  &  4.21  &  0.56 \\
$\mathrm{N_BRb_B}$    &  -1  &  0.5  &  0.57  &  10.26  &  0.54 \\
$\mathrm{O_BO_B}$    &  -1  &  1.5  &  0.65  &  14.35  &  0.3 \\
$\mathrm{Int_OO_B}$    &  1  &  1.0  &  2.99  &  4.86  &  0.72 \\
$\mathrm{B_NLi_B}$    &  -1  &  0.5  &  0.8  &  7.08  &  0.44 \\
$\mathrm{S_BV_B}$    &  1  &  0.5  &  1.09  &  6.82  &  0.5 \\
$\mathrm{Ge_NInt_{Ge}}$ &  -1  &  1.0  &  1.19  &  6.57  &  0.79 \\
$\mathrm{Al_NV_B}$    &  -1  &  1.0  &  3.64  &  6.43  &  0.79 \\
$\mathrm{Tl_NTl_N}$    &  1  &  0.5  &  0.52  &  4.77  &  0.97 \\
$\mathrm{Int_BSr_B}$    &  1  &  0.5  &  1.85  &  9.2  &  0.94 \\
$\mathrm{Si_BV_B}$   &  0  &  1.0  &  0.63  &  3.81  &  0.54 \\
$\mathrm{Be_BInt_B}$    &  -1  &  0.5  &  2.29  &  5.62  &  0.8 \\
$\mathrm{I_NV_B}$    &  0  &  0.5  &  1.5  &  3.53  &  0.92 \\
$\mathrm{B_NMg_B}$    &  0  &  0.5  &  0.72  &  5.29  &  0.44 \\
$\mathrm{Ga_BV_B}$   &  -1  &  1.0  &  0.62  &  7.5  &  0.84 \\
$\mathrm{Rb_N}$    &  1  &  0.5  &  0.51  &  11.32  &  0.53 \\
$\mathrm{Ba_BV_B}$   &  0  &  1.0  &  0.57  &  9.0  &  0.48 \\
$\mathrm{Al_BV_B}$    &  -1  &  1.0  &  0.53  &  7.37  &  0.66 \\
\end{tabular}
\end{minipage}
\label{tab:screened}
\end{table*}

\begin{table*}[htb]
\caption{Selected defects in c-BN simulated with the HSE functional using $\alpha=25\%$. Results from calculations using the PBE functional are also shown for comparison. $\Delta \mathrm{ZPL}$ is the difference in ZPL between the HSE and PBE functionals. Ground state (GS) point group symmetry is also shown.}
\begin{threeparttable}
\setlength{\tabcolsep}{4pt}
\begin{tabular} {c|cc|ccccc|cccc}
 & & & \multicolumn{5}{c|}{HSE Results} & \multicolumn{4}{c}{PBE Results} \\
Defect  & Ch & Spin & ZPL  & $\Delta$ZPL & TDM  & $\Delta$Q & GS Symmetry & ZPL & TDM  & $\Delta$Q & Binding Energy \\
 &  &  & eV & eV & Debye & amu$^{1/2}$Å & & eV & Debye & amu$^{1/2}$Å & eV\\
\hline
$\mathrm{O_NV_B}$  &  0  &  1.0  & 1.333 & 0.233 & 11.49 & 0.16 & $\mathrm{C_{3v}}$ &  1.10  &  5.07  &  0.39 & 5.35\\
%\phantom{1} $O_NV_B$ $^a$ &  0  &  1.0  & 1.540 & 0.439 $^b$ & 9.53 & 0.15 & $\mathrm{C_{3v}}$ &    &    &   &   \\
$\mathrm{O_NV_B}$  &  -1  &  0.5   & 1.448 & 0.316 & 9.94 & 0.16 & $\mathrm{C_{1h}}$ &  1.13  &  9.43  &  0.46 & 5.21\\
$\mathrm{Ca_NV_B}$ &  0  &  0.0   & 2.051 & 0.161 & 8.77 & 1.85 & $\mathrm{C_{3v}}$ &  1.89  &  8.07  &  1.63 & 7.87\\
$\mathrm{Na_B}$       &  -1  &  0.5   & 1.298 & 0.308 & 5.68 & 0.77 & $\mathrm{C_{3v}}$ &  0.99  &  7.83  &  0.14 & 12.71\\
%$C_NV_B$  &  -1  &  -0.5   & 0.800 $^c$ & -0.091 & 2.94 & 0.62 & $\mathrm{C_{1h}}$ &  0.89  &  3.5  &  0.43 & 1.93\\
 $\mathrm{C_NV_B}$ \tnote{a} &  -1  &  0.5       & 1.027    & 0.136 & 5.18 & 0.67 & $\mathrm{C_{1h}}$ &0.89  &  3.5  &  0.43 & 1.93\\
$\mathrm{C_BV_N}$  &  1  &  0.5  & 1.740 & 0.175 & 5.53 & 1.42 & $\mathrm{C_{3v}}$ &  1.56  &  4.51  &  1.25 & 1.86\\
\end{tabular}

\begin{tablenotes}\footnotesize
\item[a] HSE functional results are for the third lowest excitation, the lowest excitation in the spin up channel. See Sec~\ref{sec:carbon} for further details.
\end{tablenotes}

\end{threeparttable}
\label{tab:deltaZPL}
\end{table*}

The following defects are selected for more accurate characterization using hybrid functional. $\mathrm{O_NV_B}$ and $\mathrm{O_NV_B^-}$ passed the filter and are of interest due to earlier studies, see Table~\ref{tab:knowntheory}.
Carbon is of interest since it is a common impurity in c-BN \cite{PhysRevB.59.5233}. The $\mathrm{C_NV_B^-}$ defect passed the filter so it is considered for hybrid calculations. $\mathrm{C_BV_N^+}$ is selected due to its similarity to $\mathrm{C_NV_B^-}$. The $\mathrm{C_BV_B}$ defect is also considered in Appendix \ref{sec:app}, despite it did not pass the filter, as it was considered previously in Ref.~\cite{PhysRevB.108.L041102}. $\mathrm{Ca_NV_B}$ is selected despite high $\Delta Q$ because of the high TDM and ZPL in a region with observed lines \cite{Lopez-Morales:20}. $\mathrm{Na_B^-}$ passed the filter and had high TDM and low $\Delta Q$. The results of high accuracy calculations for these defects are summarized in Table~\ref{tab:deltaZPL} and discussed in Sec.~\ref{sec:res}.

\section{Results and Discussion}
\label{sec:res}
For the high accuracy DFT calculations, the HSE06 \cite{HSE, HSEerrata} functional is used with exact exchange mixing parameter $\alpha= 25\%$. The same lattice parameter as before (3.6246 Å) is used for the HSE functional calculations. The plane wave cutoff energy is 520 eV. The stopping parameters are $1\cdot10^{-6}$ eV for electronic relaxation and $1 \cdot 10^{-4}$ eV for ionic relaxation. As VASP 5.4.4 is used with the flags LDIAG = FALSE and ALGO = Damped, we use a version of VASP where the behavior regarding subspace rotations has been corrected, see Appendix in Ref.~\cite{PhysRevApplied.22.034056}.

VESTA is used to visualize the crystal structures \cite{Momma:ko5060}, boron is displayed as green spheres, nitrogen is gray, oxygen is red, calcium is blue, carbon is brown, and sodium is yellow. ADAQ-SYM \cite{STENLUND2025109468} is used for symmetry analysis and for generating energy level diagrams. The symmetry allowed transitions are shown with colored arrows, transitions allowed for light polarized perpendicular (parallel) to the principle axis are shown in red (blue).

\subsection{Oxygen-Vacancy Center}

\begin{figure}[]
  \includegraphics[width=\columnwidth]{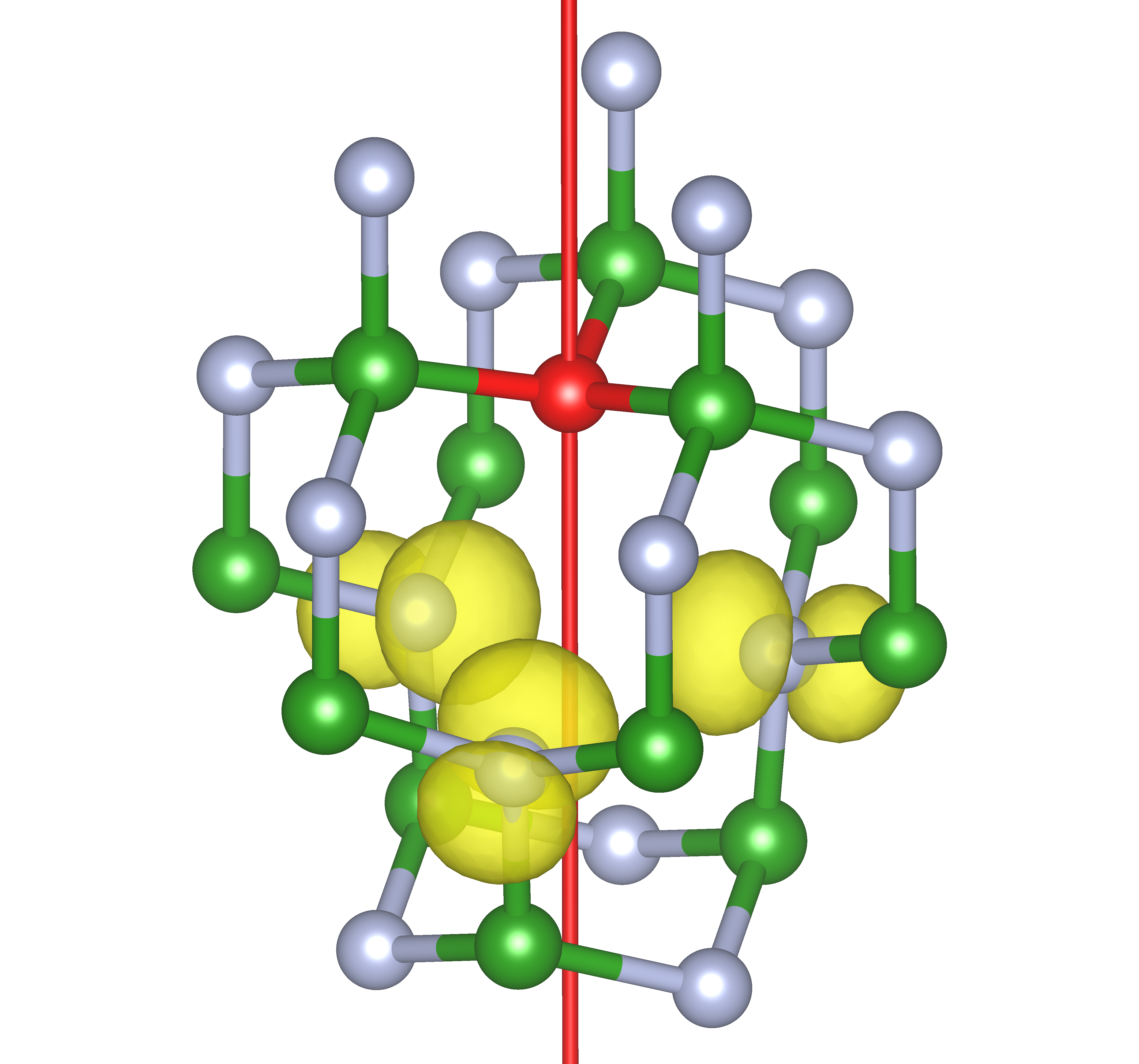}
	\caption{Crystal structure of the $\mathrm{O_NV_B}$ defect. The spin density of the neutral ground state is show with a yellow isosurface indicating a positive value (spin-up). The principle axis (1,1,1) is shown in red.}
	\label{fig:onvb_geo} 
\end{figure}

The neutral $\mathrm{O_NV_B}$ defect is similar to the $\mathrm{NV^-}$ center in diamond. The defect consists of an oxygen atom substituting a nitrogen and an adjacent boron vacancy. Figure~\ref{fig:onvb_geo} shows its crystal structure, which has $\mathrm{C_{3v}}$ symmetry. The spin density of the neutral ground state is also shown as an isosurface. Formation energy for this defect, calculated with the PBE functional, is shown in Figure~\ref{fig:onvb_form} and has been discussed in Sec.~\ref{sec:databasescreen}.

The simulated band gap, when using a 512 atom supercell and the equilibrium lattice parameter calculated with the PBE functional, is 5.8 eV when using the HSE functional with 25\% mixing and 6.2 eV with 33\% mixing, the latter is much closer to the experimental gap of 6.4 eV \cite{CHRENKO1974511}. Thus it is worth considering the effect of the functional on the ZPL. Table~\ref{tab:onvb_zpl} shows the ZPL of the $\mathrm{O_NV_B}$ defect calculated with different settings. ZPL increases with increased exact exchange parameter $\alpha$, and decreases a little with increased supercell size. The excited states are simulated without symmetry constraints. The three different lattice parameters in the table are the equilibrium lattice parameters of c-BN calculated in the PBE, HSE with $\alpha=25\%$ and $\alpha=33\%$ functionals, respectively.

\begin{figure}%
    \centering
    \includegraphics[width=0.9\linewidth]{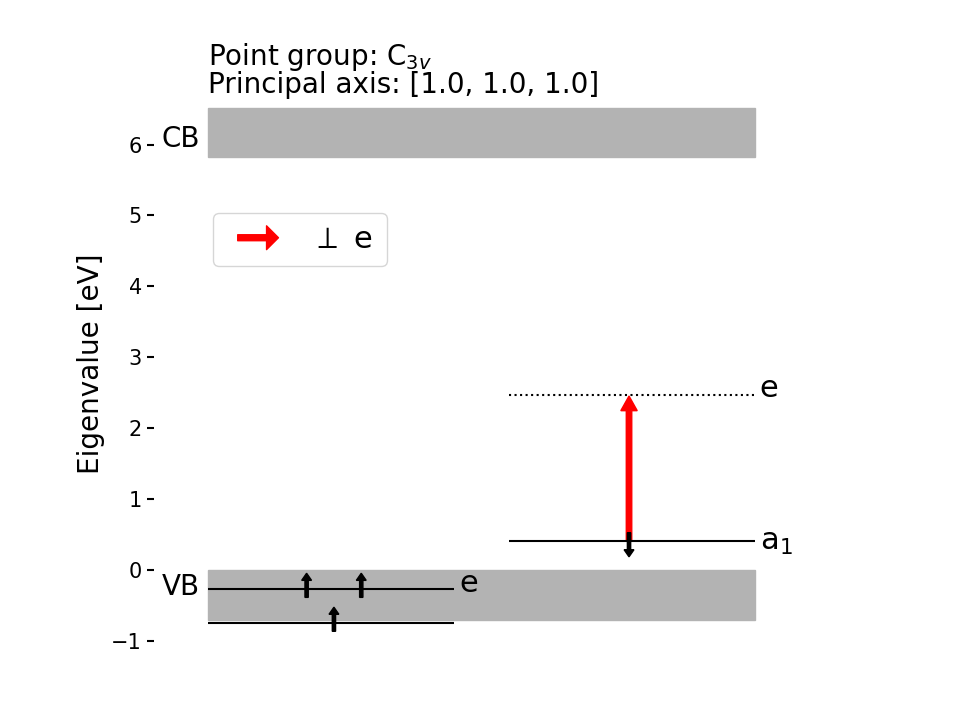}\\
    \centering a) Ground state.
    \qquad
    \includegraphics[width=0.9\linewidth]{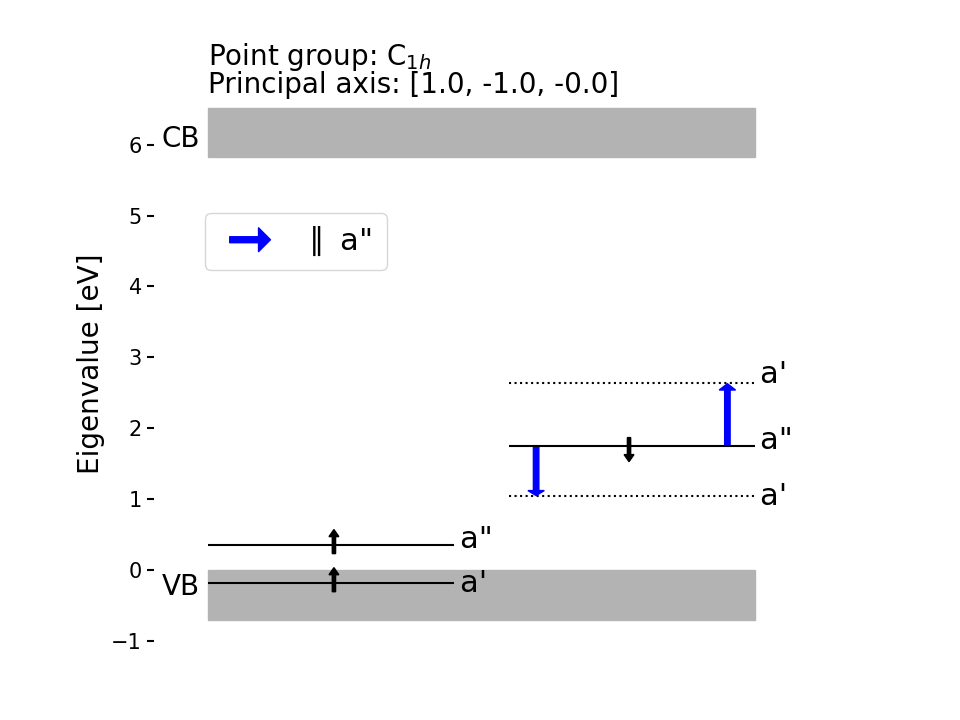}\\
    \centering b) Excited state where Jahn-Teller distortion has occured, resulting in a low symmetry geometry. 
    \includegraphics[width=0.9\linewidth]{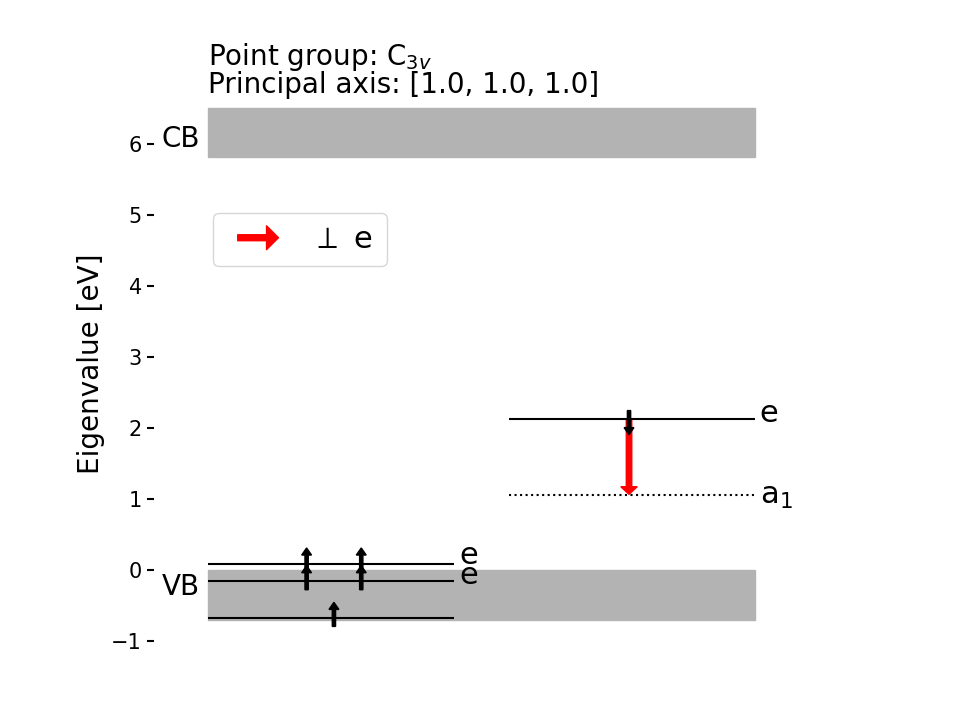}\\
    \centering c) Excited state where the $C_{3v}$ symmetry is maintained.
    \caption{Electronic structure of the neutral $\mathrm{O_NV_B}$ defect. For the low symmetry excited state shown in b) the ZPL is 1.333 eV. For the high symmetry excited state show in c) the ZPL is 1.542 eV.}%
    \label{fig:onvb_tr_neutral}%
\end{figure}

Figure~\ref{fig:onvb_tr_neutral} shows the eigenvalue levels of the neutral $\mathrm{O_NV_B}$ defect, a) shows the ground state where the only allowed transition is from an $a'$ state to an $e$ state. Relaxing the excited state without symmetry constraints results in a Jahn-Teller distortion which lowers the symmetry from $\mathrm{C_{3v}}$ to $\mathrm{C_{1h}}$ and splits the degenerate $e$ state, see Figure~\ref{fig:onvb_tr_neutral}b). For this low symmetry excitation the ZPL is 1.332 eV and the Debye-Waller factor is 13\%.
Figure~\ref{fig:onvb_tr_neutral}c) shows the excited state when simulated with symmetry constraints, the $\mathrm{C_{3v}}$ symmetry is maintained and the ZPL is 1.540 eV. This ZPL difference between high and low symmetry gives a Jahn-Teller energy ($E_{JT}$) of 0.207 eV.

\begin{figure}%
    \centering
    \subfloat[\centering Ground state ]{{\includegraphics[width=\columnwidth]{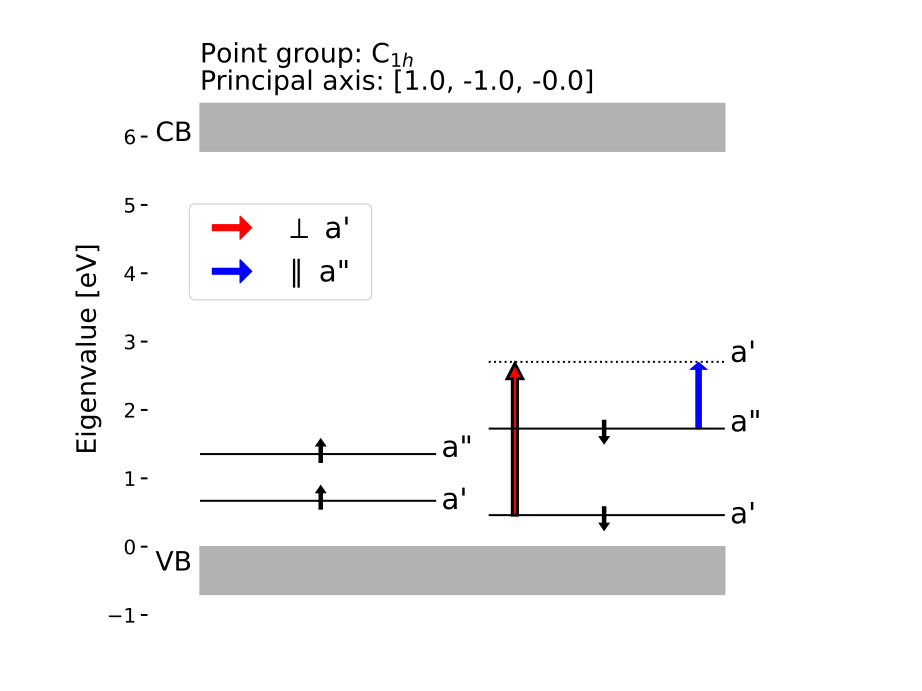} }}%
    \qquad
    \subfloat[\centering Excited state]{{\includegraphics[width=\columnwidth]{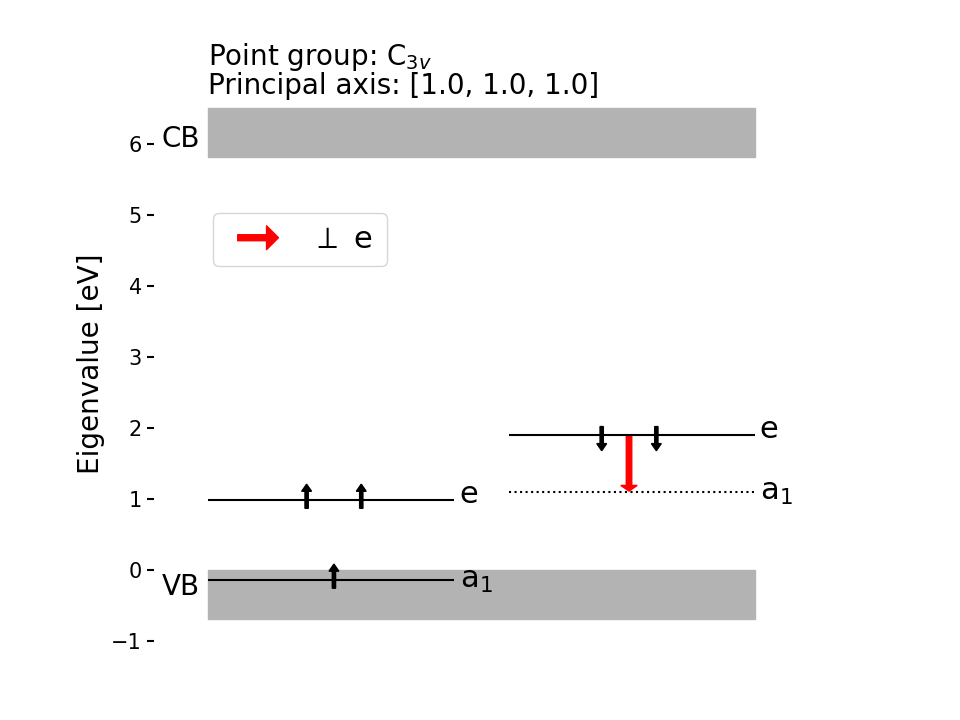} }}%
    \caption{Electronic structure of the $\mathrm{O_NV_B}$ negative charge state. a) shows the ground state, the red arrow with the black outline indicate the transition to the excited state, shown in b). The ZPL is 1.448 eV.}%
    \label{fig:onvb_tr_negative}%
\end{figure}

The negative charge state of the $\mathrm{O_NV_B}$ defect is shown in Figure~\ref{fig:onvb_tr_negative}, the ground state a) has two possible excitations either $a'$ or $a''$ to $a'$. Since the empty $a'$ state and the $a''$ state are components of a split $e$ state, promoting the $a''$ particle to $a'$ would result in the relaxing back to the ground state. Instead, the $a'$ particle is promoted to the empty $a'$ state to simulate the exited state, this results in the system relaxing to a high symmetry, $\mathrm{C_{3v}}$, configuration, seen in b). The resulting ZPL is 1.448 eV and the Debye-Waller factor is 12\%.

It has been suggested that the neutral $\mathrm{O_NV_B}$ defect may be the cause of the 1.63 eV GC-2 line observed in c-BN due to the similar ZPL and phonon sideband \cite{PhysRevLett.113.136401}. We attempted to replicate the results of Abtew et al. \cite{PhysRevLett.113.136401} and we got results that are in relatively good agreement for our high symmetry excited state, our ZPL is 1.540 eV compared to their 1.60 eV. However, the Jahn-Teller effect can distort the system into the low symmetry configuration which has lower energy and then light would be emitted at lower energy. The $E_{JT}$ of the exited state is 207 meV, which is much higher than the $\mathrm{NV^-}$ center which has an $E_{JT}$ of 42 meV \cite{PhysRevB.96.081115}. Phonon dispersion calculations suggests a highest phonon energy at around 160 meV \cite{PhysRevApplied.20.034055}. Due to the large $E_{JT}$ it is not likely for the high symmetry excited state to be populated since a two-phonon process would be needed. 

The $\mathrm{O_NV_B^-}$ defect has a ZPL of 1.448 eV for the case when the excited state geometry and electronic structure has been relaxed without symmetry constraints. This is around 0.18 meV away from the GC-2 line. As a follow up, the $\mathrm{O_NV_B^-}$ defect was simulated with the HSE functional using a exact exchange mixing parameter $\alpha=33\%$, resulting in a ZPL of 1.600 eV. These results make $\mathrm{O_NV_B^-}$ a more plausible defect candidate as the source for the GC-2 line. 

The GC-2 line has been observed in several experiments \cite{Tkachev1985, Shipilo1986, PhysRevB.98.094106, doi:10.1021/acs.cgd.5c00722, doi:10.1021/acsami.4c17814}, however the connection to the $\mathrm{O_NV_B}$ defect has not yet been specifically investigated. We propose the following experiments to examine the connection of the GC-2 line and the $\mathrm{O_NV_B}$ defect; 1) Measuring the ZPLs of c-BN samples with different oxygen concentrations and seeing if there is a higher concentration of the color center in the samples with more oxygen. 2) An optically detected magnetic resonance (ODMR) experiment could measure the zero-field splitting parameters D and E, and then compare to theory \cite{10.1063/1.5083076}. 3) The PL spectrum of c-BN samples with a range of p- and n-dopings could be investigated to determine if both charge states of the $\mathrm{O_NV_B}$ defect are detected and at what wavelength, as well as attaining an estimate of the charge transition level. This last experiment may be challenging as controllable p- and n-doping of c-BN remains difficult \cite{PhysRevB.105.054101}, and further developments of doping methods would be needed.

\begin{table}[h!]
\caption{$\mathrm{O_NV_B}$ and $\mathrm{O_NV_B^-}$ simulated at the $\Gamma$-point. Calculated ZPL is presented for supercells of 216 or 512 atoms and with different lattice parameters. The three different lattice parameters are the equilibrium lattice parameters for the PBE, HSE with $\alpha=25\%$ and $\alpha=33\%$ functionals,  respectively.}
\begin{threeparttable}
\begin{tabular} {c|c|c|c|c}
Defect & Functional, $\alpha$ & ZPL (eV) & Lattice param. (Å) & Atoms\\
\hline
 \multirow{6}{*}{$\mathrm{O_NV_B}$} & PBE & 1.100 & 3.6246 & 512\\
 & HSE, $25\%$ & 1.332 & 3.6246 & 512 \\
 & HSE, $25\%$ & \phantom{} 1.540 \tnote{a} & 3.6246 & 512 \\
 & HSE, $25\%$ & 1.399 & 3.5979 & 512  \\
 & HSE, $25\%$ & 1.422 & 3.5979 & 216 \\
 & HSE, $33\%$ & 1.478 & 3.5898 & 512 \\
 %\hline
 & HSE, \tnote{b} & \phantom{} 1.60 \tnote{b} & Unknown & 216 \\
 \hline
 \multirow{3}{*}{$\mathrm{O_NV_B^-}$} & PBE & 1.132 & 3.6246 & 512 \\
 & HSE, $25\%$ & 1.448 & 3.6246 & 512 \\
 & HSE, $33\%$ & 1.600 & 3.6246 & 512 \\
 \hline
\end{tabular}

\begin{tablenotes}\footnotesize
\item[a] Excited state is relaxed with symmetry constraints, resulting in $\mathrm{C_{3v}}$ symmetry.
\item[b] HSE functional exact exchange mixing paramenter $\alpha$ is not specified in Ref.~\cite{PhysRevLett.113.136401}.
\end{tablenotes}

\end{threeparttable}
\label{tab:onvb_zpl}
\end{table}

\subsection{Calcium Defect}
\begin{figure}[]
  \includegraphics[width=\columnwidth]{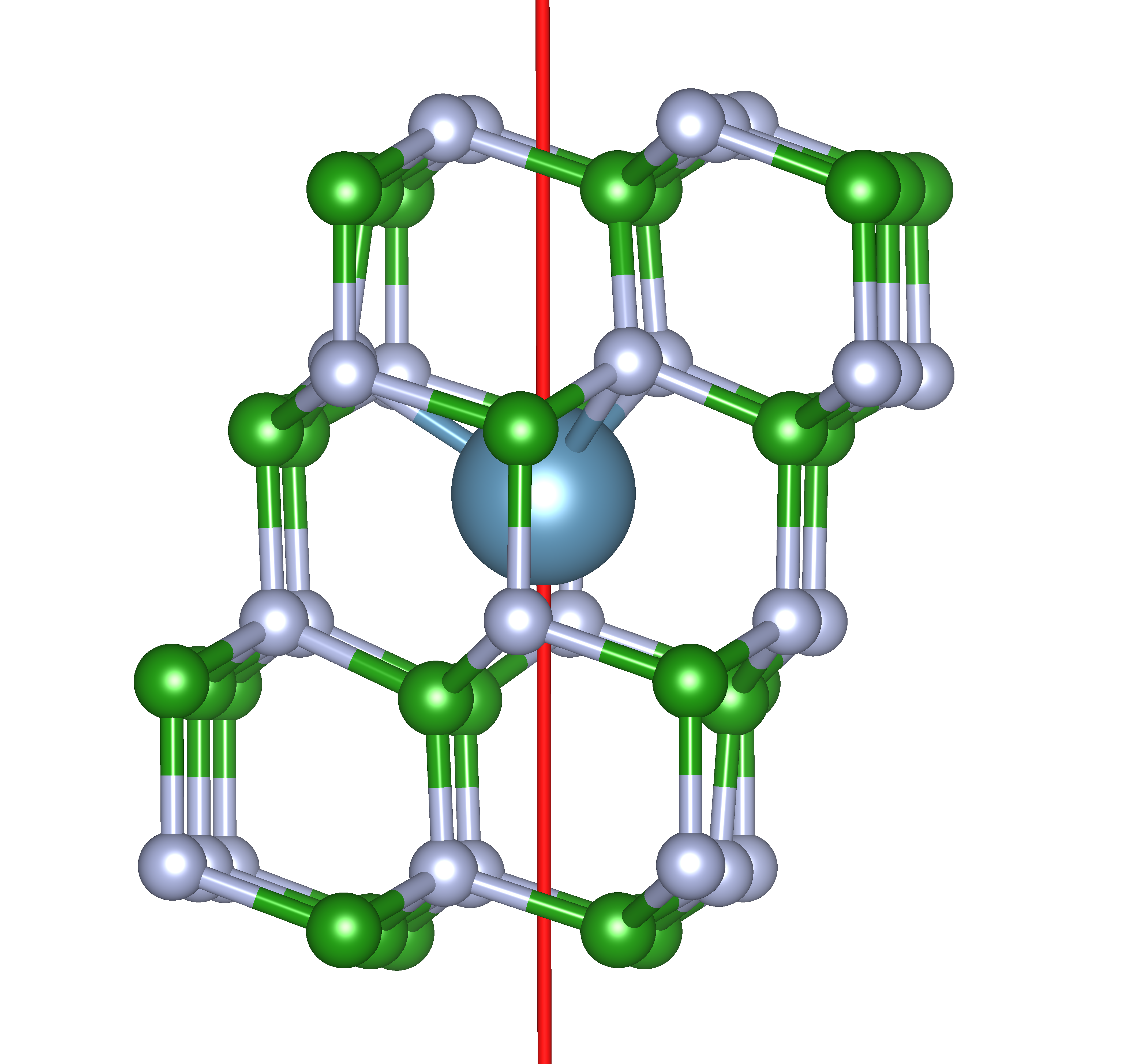}
	\caption{Crystal structure of the $\mathrm{Ca_NV_B}$ defect. The (1, -1, -1) symmetry axis is marked in red. }
	\label{fig:canvb_geo} 
\end{figure}

\begin{figure}[]
    \centering
    \subfloat[\centering Ground state]{{\includegraphics[width=\columnwidth]{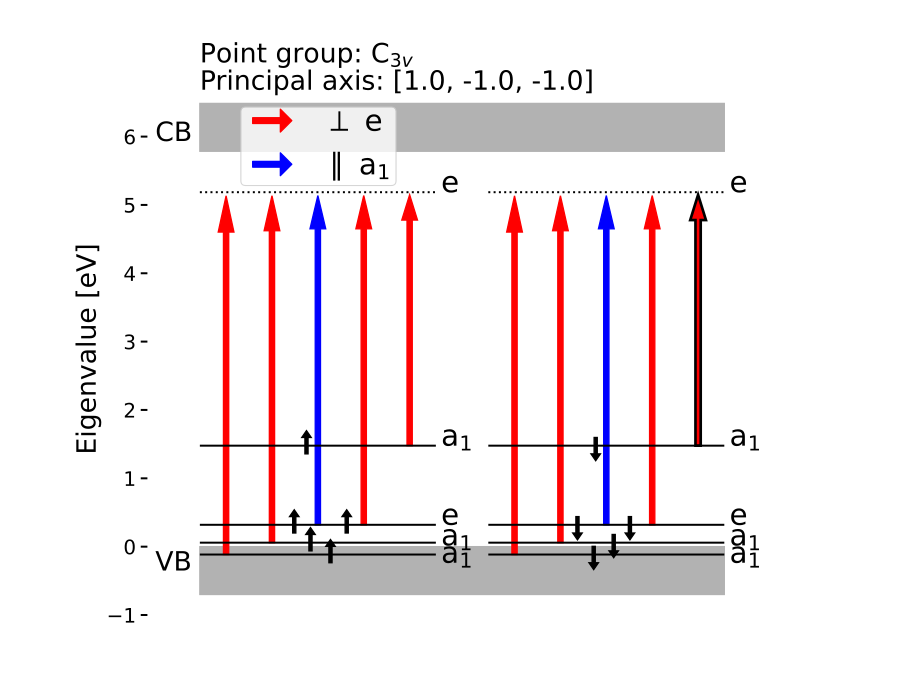} }}%
    \qquad
    \subfloat[\centering Excited state]{{\includegraphics[width=\columnwidth]{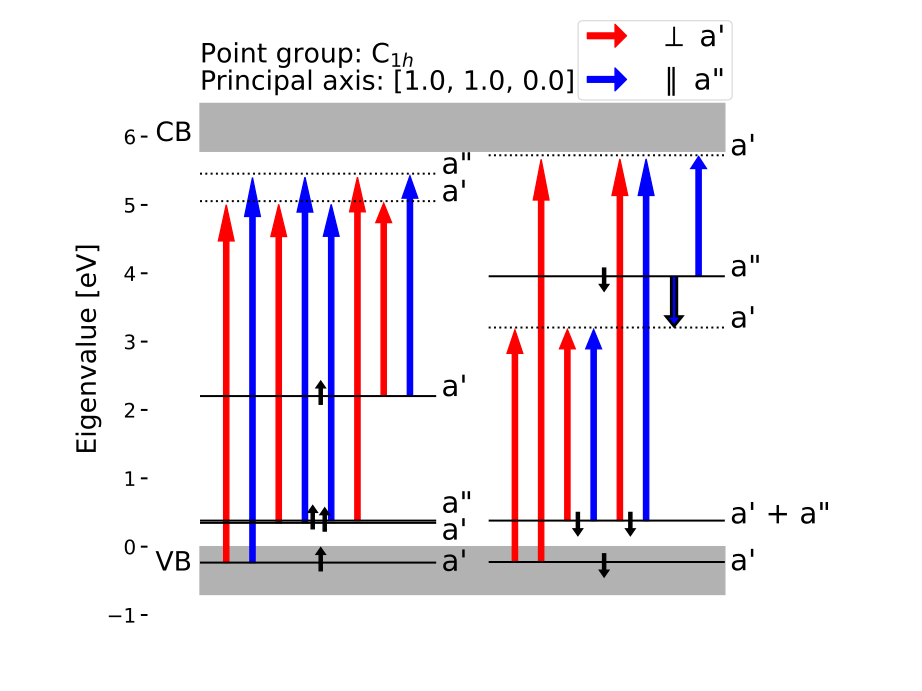} }}%
	\caption{Electronic structure of the $\mathrm{Ca_NV_B}$ defect. The arrows corresponding to the transition are are marked with a black outline for clarity. The ZPL is 2.051 eV.}
	\label{fig:canvb_transitions} 
\end{figure}

Photoluminescence lines emitting at around 500-650 nm have been found in experiment \cite{PhysRevB.98.094106, Lopez-Morales:20}, so defects with ZPLs 1.9-2.5 eV are of interest. The $\mathrm{Ca_NV_B}$ defect has a ZPL in this range. The defect is a split vacancy defect with a geometry similar to the SiV split vacancy center in diamond \cite{PhysRevB.88.235205}. Figure~\ref{fig:canvb_geo} shows the geometry of the defect, where it can be seen that the Ca atom has relaxed to the center between the N atom it replaced and the nearby B vacancy. The defect has $\mathrm{C_{3v}}$ symmetry and the principle axis, which is one of the body-diagonals of the cubic supercell, is shown in red in Figure~\ref{fig:canvb_geo}. $\mathrm{Ca_NV_B}$ has a somewhat high formation energy of 10.67 eV, but it does have a positive binding energy, meaning that $\mathrm{Ca_N}$ and $\mathrm{V_B}$ defects lower their formation energy when combining to a $\mathrm{Ca_NV_B}$ center. 
Figure~\ref{fig:canvb_transitions} shows the allowed transitions of the defect.
Since it is spin-0, there is no difference in the excitations between the spin channels. The lowest excitation is an excitation from an $a'$ state to an $e$ state, which is only allowed for light polarized perpendicularly to the principle axis. The downsides of this defect is zero spin and the high $\Delta$Q, resulting in a Debye-Waller factor of $2\cdot10^{-13}$ in the one-phonon approximation. However, it has a high TDM of 8.07 Debye, suggesting a bright defect and the ZPL is 2.051 eV.  This is in the region where there are experimentally observed photoluminescence lines without attribution \cite{Lopez-Morales:20}, although it is unlikely that the observed lines are caused by calcium defects because of the very low Debye-Waller factor. Moreover, there is no indication that calcium is an impurity in the samples in Ref.~\cite{Lopez-Morales:20}.

\subsection{Carbon Defects}
\label{sec:carbon}

\begin{figure}[h!]
  \includegraphics[width=\columnwidth]{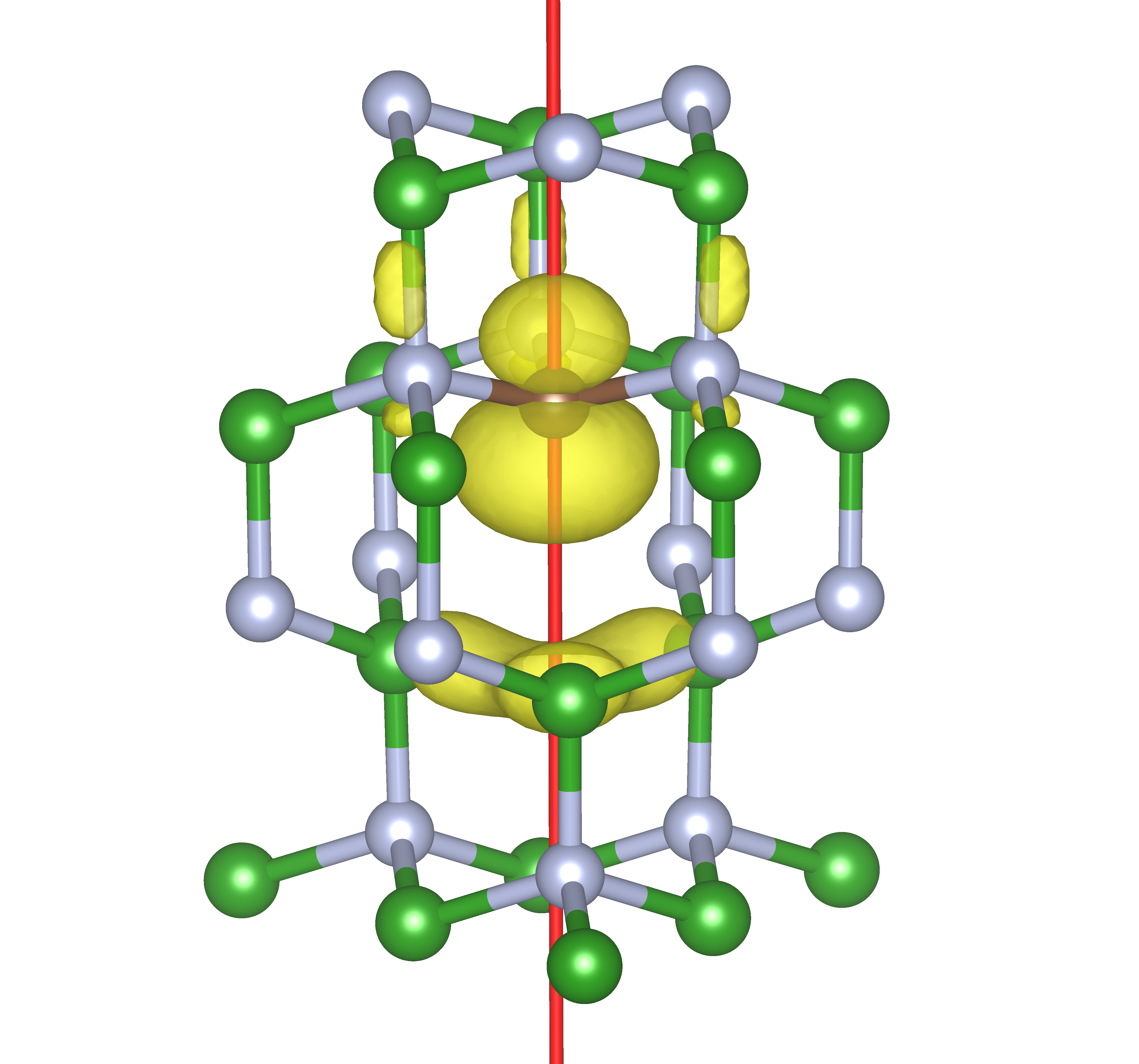}
	\caption{Crystal structure of the $\mathrm{C_BV_N^+}$ defect. The (1, 1, -1) symmetry axis is marked in red. The spin
density of the ground state is is show with a yellow isosurface indicating a positive value (spin-up).}
	\label{fig:cbvn_geo} 
\end{figure}

\begin{figure}[h!]
  \includegraphics[width=\columnwidth]{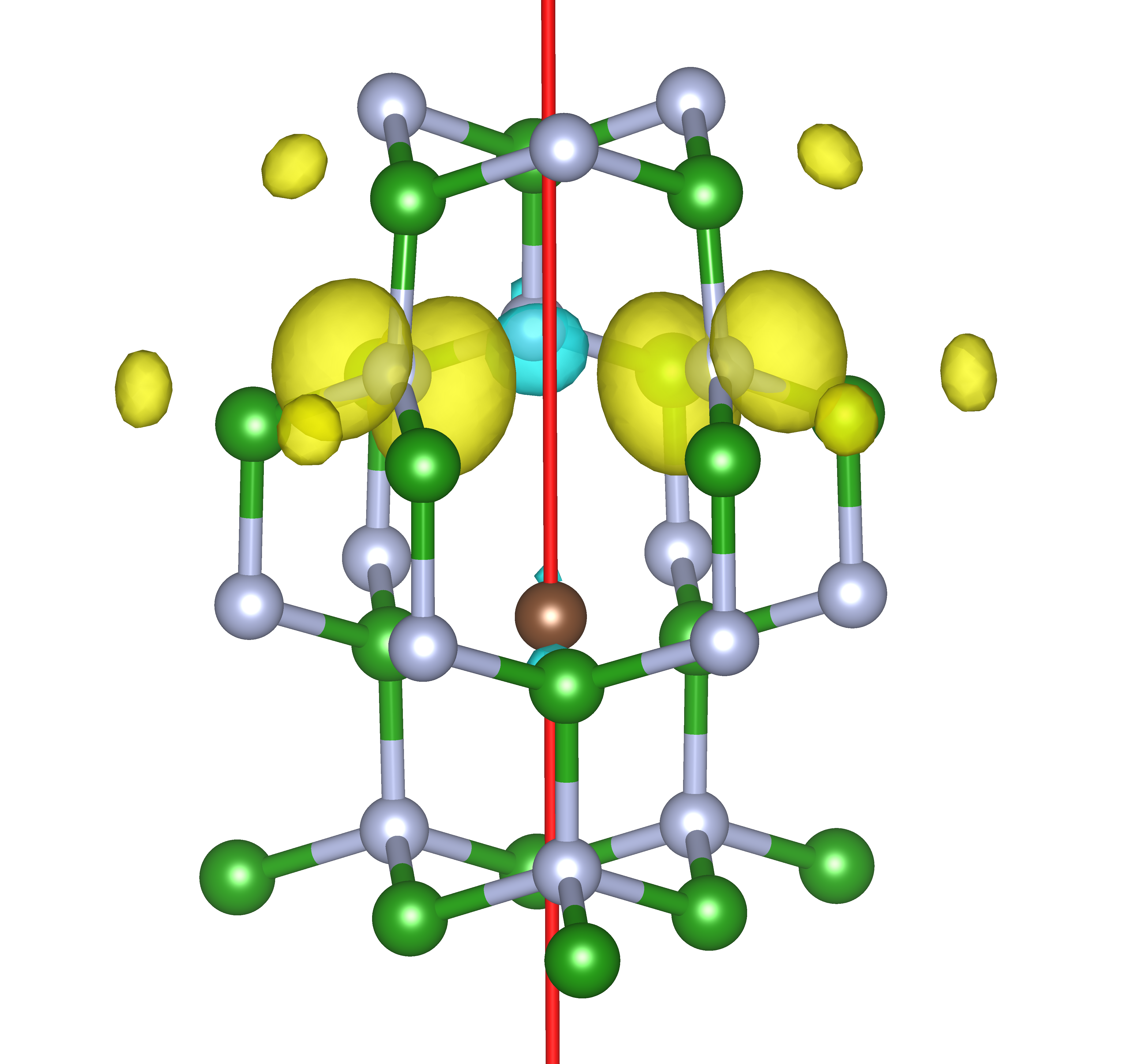}
	\caption{Crystal structure of the $\mathrm{C_NV_B^-}$ defect. The (1, -1, -1) axis is marked in red, this is the principle axis for the excited state with $\mathrm{C_{3v}}$ symmetry. This symmetry is broken in the ground state, reduced to $\mathrm{C_{1h}}$. The spin density of the ground state is is show with a yellow (spin-up) and blue (spin-down) isosurface.}
	\label{fig:cnvb_geo} 
\end{figure}

Carbon is a commonly occurring impurity in boron nitrides \cite{doi:10.1021/acs.nanolett.9b02879, PhysRevMaterials.6.014005} and there are carbon based point defects investigated in h-BN \cite{Mendelson2021, Babar2025}. Here, two carbon defects in c-BN are presented, the $\mathrm{C_BV_N^+}$ and $\mathrm{C_NV_B^-}$ defects. Their geometries are presented in Figures~\ref{fig:cbvn_geo} and \ref{fig:cnvb_geo}. This pair of defects comes from a carbon atom binding to a divacancy on either the B or N site. The $\mathrm{C_BV_N^+}$ defect has $\mathrm{C_{3v}}$ symmetry, and the $\mathrm{C_NV_B^-}$ defect has $\mathrm{C_{1h}}$ symmetry due to a Jahn-Teller distortion breaking the $\mathrm{C_{3v}}$ symmetry. The principle axes of the $\mathrm{C_{3v}}$ symmetry  are shown in Figures~\ref{fig:cbvn_geo} and \ref{fig:cnvb_geo}, which are two of the four body-diagonals of the cubic supercell.

\begin{figure}[]
    \centering
    \subfloat[\centering Ground state]{{\includegraphics[width=\columnwidth]{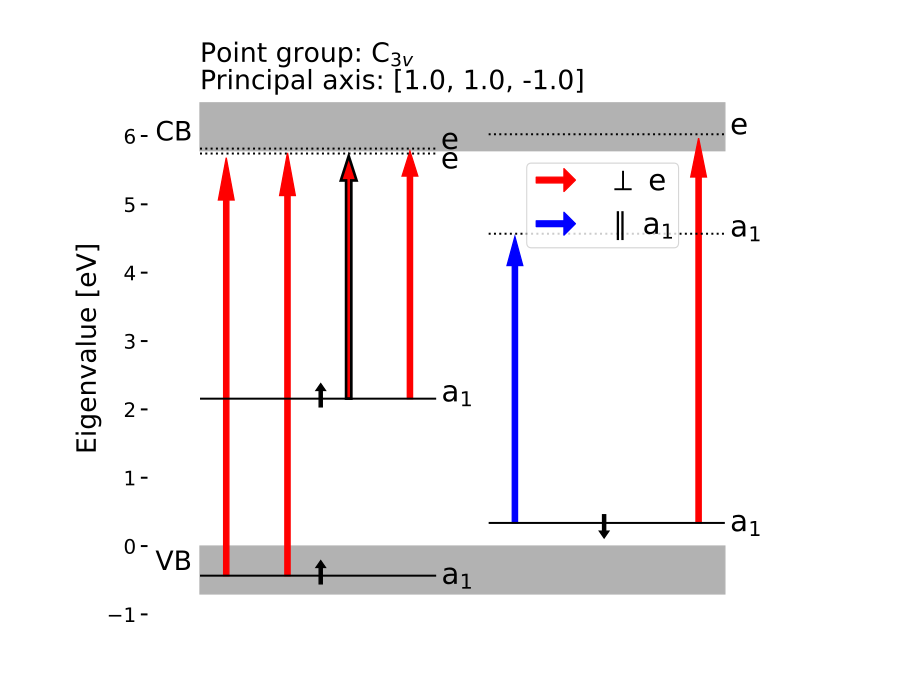} }}%
    \qquad
    \subfloat[\centering Excited state]{{\includegraphics[width=\columnwidth]{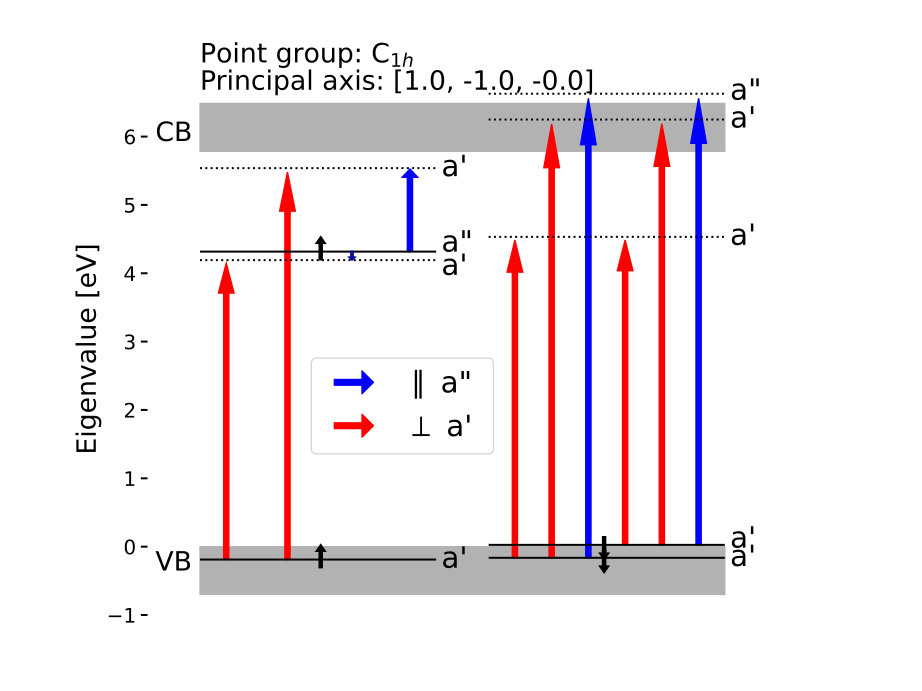} }}%
	\caption{Electronic structure of the $\mathrm{C_BV_N^+}$ defect. The ZPL is 1.740 eV.}
	\label{fig:cbvn_transitions} 
\end{figure}

The electronic structure of the the $\mathrm{C_BV_N^+}$ defect is shown in Figure~\ref{fig:cbvn_transitions}, the transition is from a $a_1$ state to a shallow $e$ state. This results in a ZPL of 1.740 eV and the TDM is 5.53 Debye. The defect has a high $\Delta$Q of 1.42 amu$^{1/2}$Å and a low Debye-Waller factor of $1\cdot10^{-9}$. This defect would be likely to exist in a c-BN crystal with carbon impurities. 

\begin{figure}[]
    \centering
    \subfloat[\centering Ground state]{{\includegraphics[width=\columnwidth]{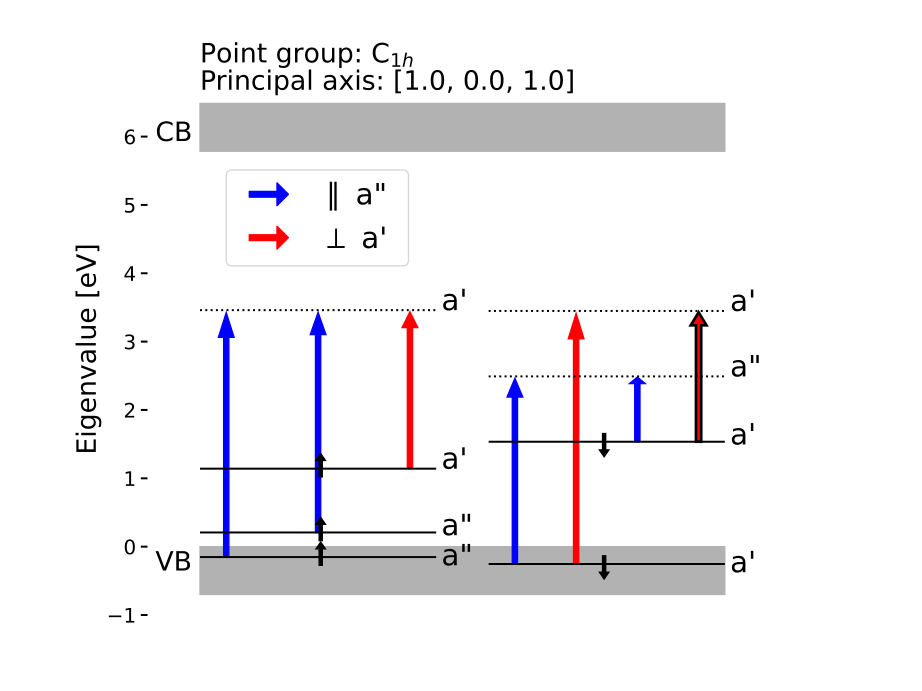} }}%
    \qquad
    \subfloat[\centering Excited state]{{\includegraphics[width=\columnwidth]{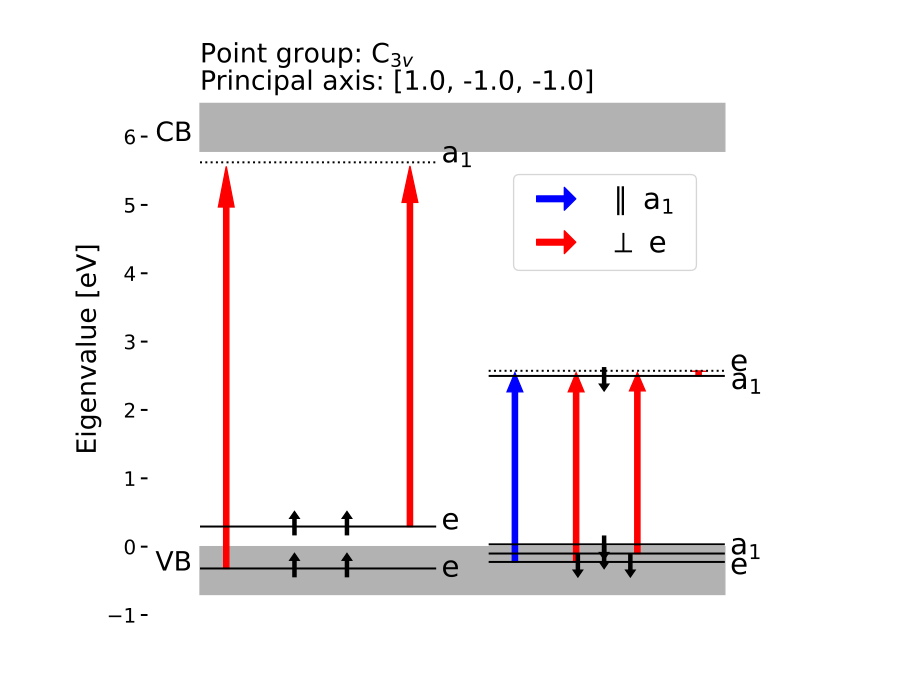} }}%
    \qquad
    \subfloat[\centering Third lowest excited state, the lowest excitation in the spin up channel]{{\includegraphics[width=\columnwidth]{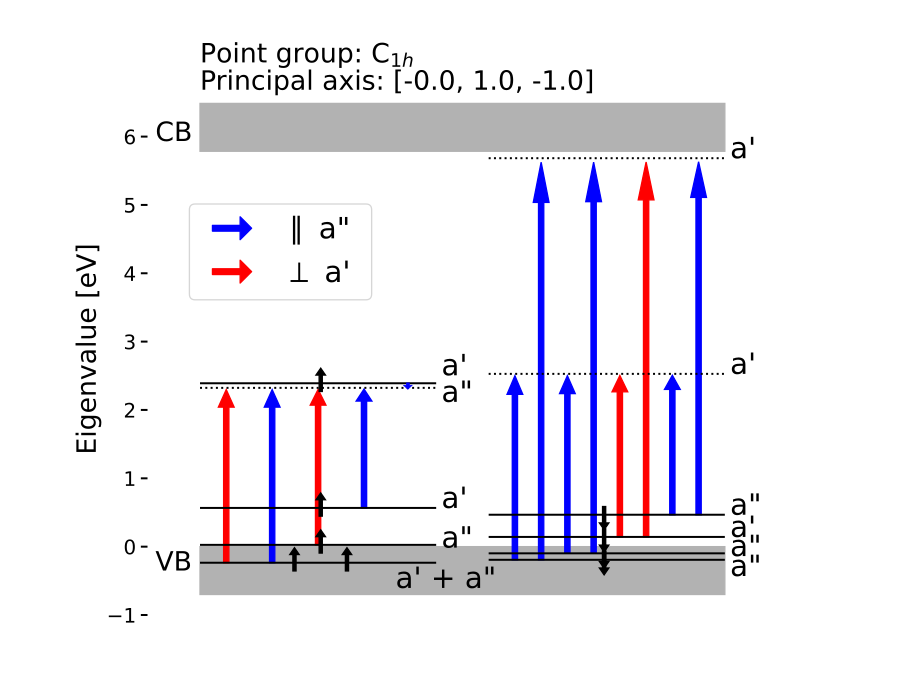} }}%
	\caption{Electronic structure of the $\mathrm{C_NV_B^-}$ defect. The ZPL of the lowest excitation is 0.800, and the second lowest is 1.027 eV.}
	\label{fig:cnvb_transitions} 
\end{figure}

For the $\mathrm{C_NV_B^-}$ defect, several excitations are investigated. In the ADAQ workflow, the empty $a''$ state is too close to the occupied $a'$ state, thus the second lowest excitation is simulated, giving a ZPL of 0.89 eV with the PBE functional. 
The lowest excitation in the spin down channel, which ADAQ skipped, relaxes back to the ground state when simulated with the HSE functional. When simulating this defect with the HSE functional, there is a greater splitting of these states, shown in Figure~\ref{fig:cnvb_transitions}a), and when relaxing the geometry in the second lowest excitation, there is a state switching where the occupied orbital has a lower eigenvalue than the empty orbital, see Figure~\ref{fig:cnvb_transitions}b), the resulting ZPL is 0.800 eV. 

Another excitation in the spin up channel is also simulated, shown in Figure~\ref{fig:cnvb_transitions}c). This results in a ZPL of 1.027 eV, and the order of the orbital eigenvalues do not switch. The TDM is 5.18 Debye, the $\Delta$Q is 0.67 amu$^{1/2}$Å and the Debye-Waller factor is $6\cdot10^{-4}$. Thus,  $\mathrm{C_NV_B^-}$ is a potential emitter close to the edge of the telecom O-band.

There are other carbon defects that may be of interest, the $\mathrm{C_BV_B}$ defect (also $\mathrm{Si_BV_B}$) that has been investigated previously \cite{PhysRevB.108.L041102, NGUYEN2025113388}, appear in our database. $\mathrm{C_BV_B}$ did not pass filter, but it is discussed further in Appendix~\ref{sec:app}.
Another defect that may be of interest is $\mathrm{C_NV_N^+}$. From the calculation in ADAQ, the $\mathrm{C_NV_N^+}$ defect has a ZPL of 0.66 eV and a high TDM of 14.74 Debye, see Table~\ref{tab:screened}. However, the region of stability for the positive charge state is very narrow, the positive to neutral charge transition occurs at a Fermi energy of 0.61 eV. Heavy p-doping may be required to actually observe this state.

\subsection{Sodium Defect}

\begin{figure}[]
  \includegraphics[width=\columnwidth]{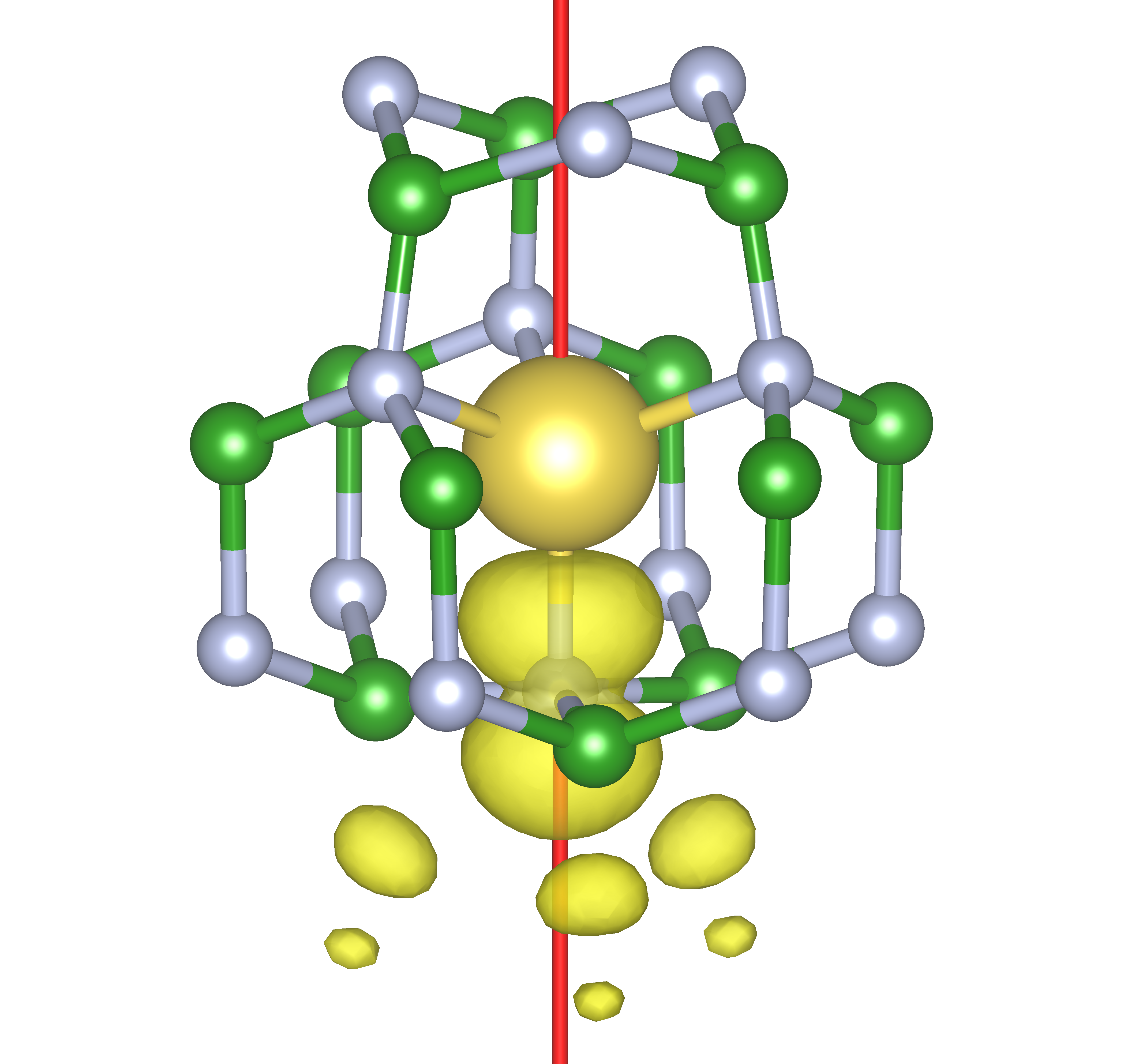}
	\caption{Crystal structure of the $\mathrm{Na_B^-}$ defect. The principle axis (1,1,1) is shown in red and the spin density of the ground state is is show with a yellow isosurface indicating a positive value (spin-up).}
	\label{fig:nab_geo} 
\end{figure}

\begin{figure}[]
    \centering
    \subfloat[\centering Ground state]{{\includegraphics[width=\columnwidth]{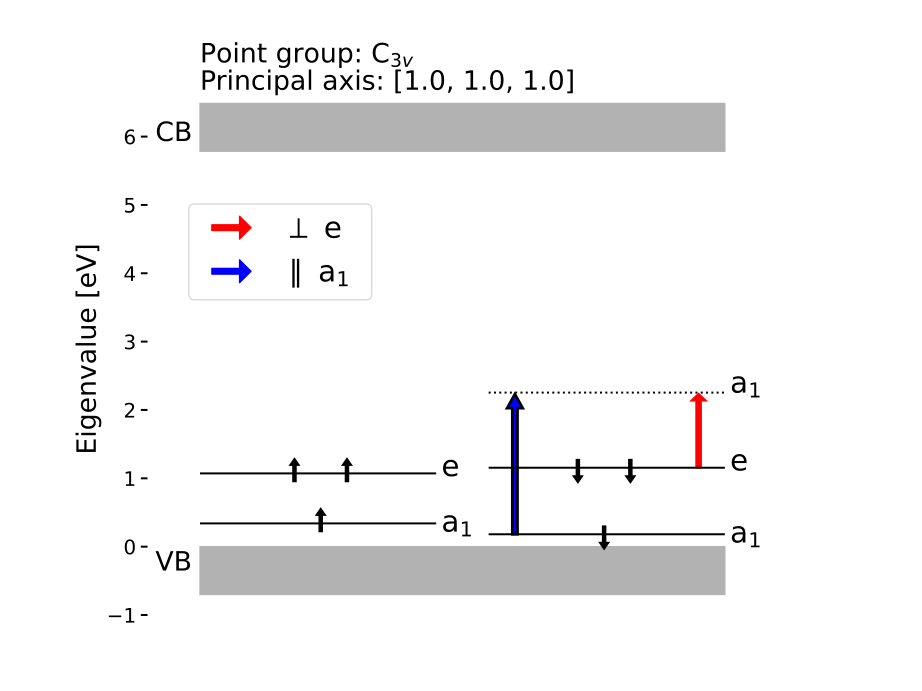} }}%
    \qquad
    \subfloat[\centering Excited state]{{\includegraphics[width=\columnwidth]{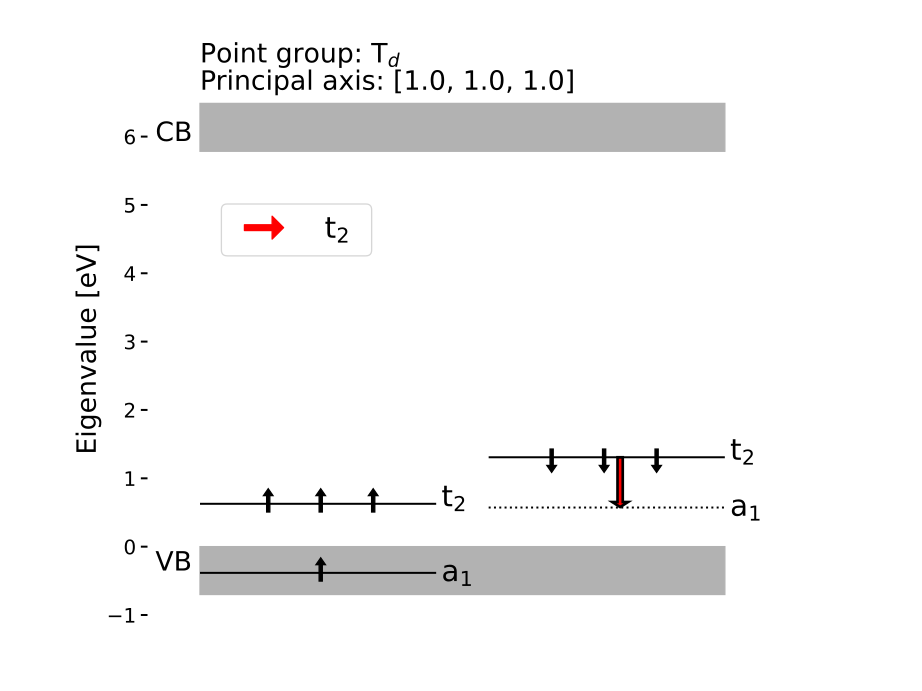} }}%
	\caption{Electronic structure of the $\mathrm{Na_B^-}$ defect. The ZPL is 1.298 eV.}
	\label{fig:nab_transitions} 
\end{figure}

The sodium substitutional defect is a new defect identified via the ADAQ workflow. Figure~\ref{fig:nab_geo} shows the defect geometry. It is a substitutional defect where a Jahn-Teller (JT) distortion lowers the symmetry from $\mathrm{T_{d}}$ to $\mathrm{C_{3v}}$. The electronic structure and the two possible excitations are shown in Figure~\ref{fig:nab_transitions}. The three uppermost orbitals, $a_1$ and $e$, are part of a triply degenerate state that is split by the JT distortion. Hence, the relevant excitation is from the occupied $a'$ state to the empty $a_1$ state. The ZPL is 1.298 eV, the TDM is 5.68 Debye, and the Debye-Waller factor is $7\cdot10^{-5}$, as calculated with the HSE functional. When simulated with the PBE functional, the JT distortion distortion is small, but with the HSE functional the distortion becomes much larger and the empty $a_1$ and occupied $e$ states split more than in the simulation using the PBE functional. These properties indicate that the $\mathrm{Na_B^-}$ defect is a potentially bright emitter in the near-infrared range, although it would have a large phonon sideband. 
The $\mathrm{Na_B^-}$ defect is similar to the sodium substitutional defect in diamond, also discovered with ADAQ \cite{Davidsson2024}.

\newpage

\section{Conclusion}
\label{sec:conc}
Using the ADAQ framework, we simulated a large number of defects in c-BN and characterized their stability and optical properties. The calculated properties are entered into the ADAQ database which enables the search for defects that are promising candidates for quantum technology. The results of the screening workflow suggests that c-BN hosts multiple color centers that emit light brightly and in relevant wavelength ranges, such as the telecom (1260-1675 nm) and biological (650-1350 nm) windows.
A handful of defects with promising properties from the ADAQ simulations are further investigated with high accuracy simulations using hybrid DFT calculations. The $\mathrm{C_NV_B^-}$ and $\mathrm{C_BV_N^+}$ defects are two new defects that are likely to exist in c-BN due to the prevalence of carbon. Both color centers are bright with spin-1/2 and ZPLs at 1.027 eV and 1.740 eV, respectively. However, $\mathrm{C_BV_N^+}$ has a low Debye-Waller factor meaning it may be hard to distinguish the ZPL from the phonon sideband. $\mathrm{Na_B^-}$ is a newly identified bright defect with spin-1/2 and the ZPL of 1.298 eV. $\mathrm{Ca_NV_B}$ is a bright spin-0 defect with the ZPL 2.051 eV. The previously studied $\mathrm{O_NV_B}$ defect is also captured in our screening. It is examined in several configurations, and its negative charge state is predicted to be a bright emitter. The neutral charge state has been considered and we achieve relatively good agreement with previous results for the high symmetry configuration of the excited state. However, we find that the low symmetry excited state, due to a Jahn-Teller distortion, has lower energy and it is more likely to be populated. The ZPL calculated for the low symmetry configuration of the excited state would not match the suggested GC-2 line at 1.63 eV. We suggest $\mathrm{O_NV_B^-}$ as a more plausible candidate for explaining the GC-2 line.

\section*{Data Availability}
Defect data for c-BN (and other hosts) can be found at https://defects.anyterial.se/, other data can be provided upon reasonable request.

\section*{Acknowledgments}

Many thanks to Mark Turiansky and Chris Van de Walle for discussions at conferences, and for sending the input files they used for simulating the $\mathrm{C_BV_B}$ defect in Ref. \cite{PhysRevB.108.L041102}.
This work was partially supported by the Knut and Alice Wallenberg Foundation through the Wallenberg Centre for Quantum Technology (WACQT). Support from the Swedish Government Strategic Research Area Swedish e-science Research Centre (SeRC) and the Swedish Government Strategic Research Area in Materials Science on Functional Materials at Linköping University (Faculty Grant SFO-Mat-LiU No. 2009 00971) are gratefully acknowledged. JD acknowledges support from the Swedish Research Council (VR) Grant No. 2022-00276. The computations were enabled by resources provided by the National Academic Infrastructure for Supercomputing in Sweden (NAISS) and the Swedish National Infrastructure for Computing (SNIC) at NSC, partially funded by the Swedish Research Council through grant agreements no. 2022-06725 and no. 2018-05973. This research was  supported by the National Research, Development, and Innovation Office of Hungary within grant FK 145395. This project is funded by the European Commission within Horizon Europe projects (Grant Nos. 101156088 and 101129663).

\appendix
\section{$\mathrm{C_BV_B}$}
\label{sec:app}

\renewcommand{\thefigure}{A\arabic{figure}}
\setcounter{figure}{0}

The neutral spin-1 $\mathrm{C_BV_B}$ defect is previously known \cite{PhysRevB.108.L041102}.
The ground state is present is the ADAQ database but the excited state is not. No excited state was calculated since the eigenvalue differences between the local states was too low, that is, the estimated ZPL was too low. This is due to the underestimation of band gaps and ZPLs in simulations using the PBE functional. When this excitation was simulated later with the PBE functional the resulting ZPL is 0.45 eV, too low for our filtering. 
%Problems were encountered when the $\mathrm{C_BV_B}$ defect was simulated with HSE with $\alpha=25\%$ using the method described in Section \ref{sec:res}. There were issues with the convergence of the excited state, these issues are likely related to the use of the flag LDIAG = FALSE in the ground state, this does not affect the total energy but it may affect the ordering of the orbitals, and since the excited state calculations are started from the WAVECAR of the ground state an unintended excited state was simulated. With a good starting WAVECAR, another issue occurs when the excited state was simulated with constrained occupation, the states switched order, that is, the occupied orbital had a lower eigenvalue than the empty orbital after some relaxation. This resulted in a ZPL of 0.577 eV, much lower than the 0.95 eV ZPL reported by Turiansky and Van de Walle \cite{PhysRevB.108.L041102}, to replicate their results we requested and received their input files (Thanks Mark!).

Convergence issues were encountered when the excited state of the $\mathrm{C_BV_B}$ defect was simulated with the HSE functional with $\alpha=25\%$ using the method described in Section \ref{sec:res}. We did not reproduce the ZPL reported by Turiansky and Van de Walle \cite{PhysRevB.108.L041102}. To replicate their results, we requested and received their input files.

Table~\ref{tab:cbvb_zpl} presents the results of our replication of the $\mathrm{C_BV_B}$ defect using the HSE functional with 33\% exact exchange mixing, plane wave cutoff energy of 520 eV and a 216 atom supercell. When simulating the system using the special k-point (1/4, 1/4, 1/4) the results are in very good agreement, and when using the $\Gamma$-point or a 2x2x2 k-point mesh the ZPL is only slightly larger. The $\Gamma$-point simulation is in closer agreement to the simulation with the denser mesh than the special k-point. The $\Gamma$-point simulation results in a ZPL of 1.009 eV, TDM of 7.26 Debye, $\Delta$Q of 0.35 amu$^{1/2}$Å, and Debye-Waller factor of 2.2\%. These are promising properties for a color center.

\begin{figure}[]
  \includegraphics[width=\columnwidth]{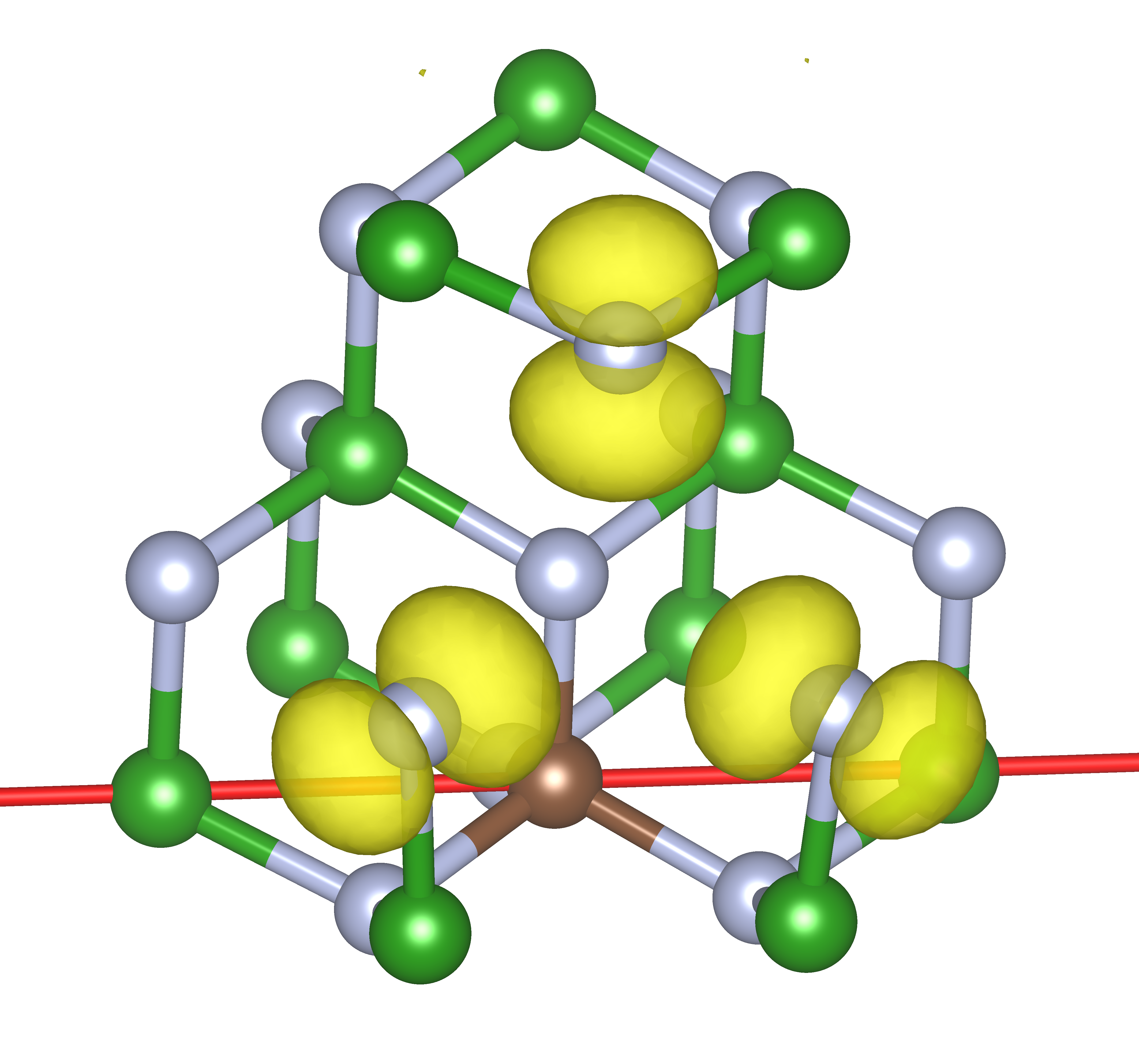}
	\caption{Crystal structure of the $\mathrm{C_BV_B}$ defect. The (1, 0, 1) symmetry axis is marked in red. The spin density of the ground state is is show with a yellow isosurface indicating a positive value (spin-up).}
	\label{fig:cbvb_geo} 
\end{figure}

\begin{figure}[]
    \centering
    \subfloat[\centering Ground state]{{\includegraphics[width=\columnwidth]{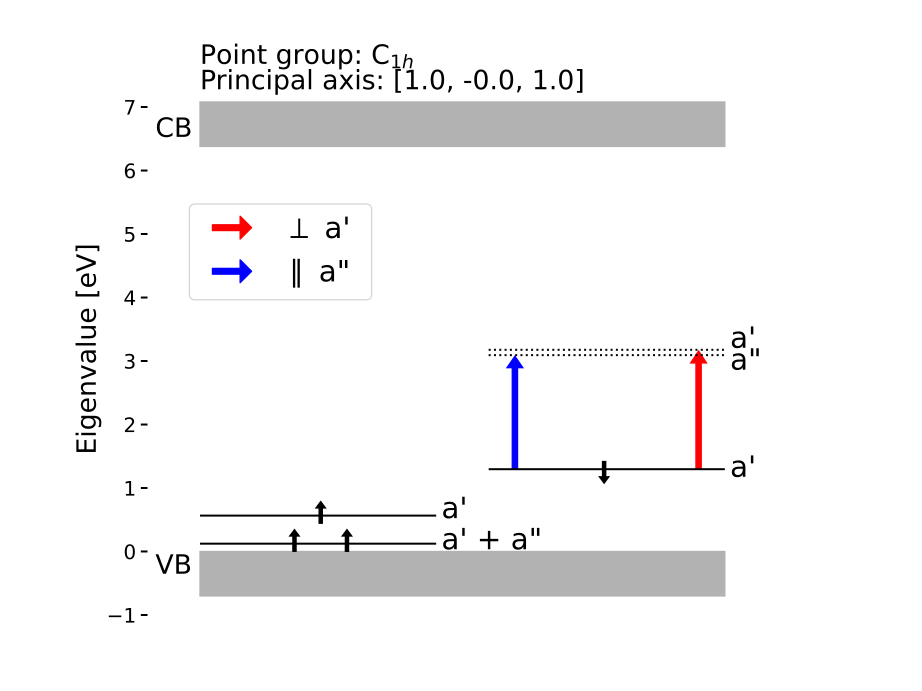} }}%
    \qquad
    \subfloat[\centering Excited state]{{\includegraphics[width=\columnwidth]{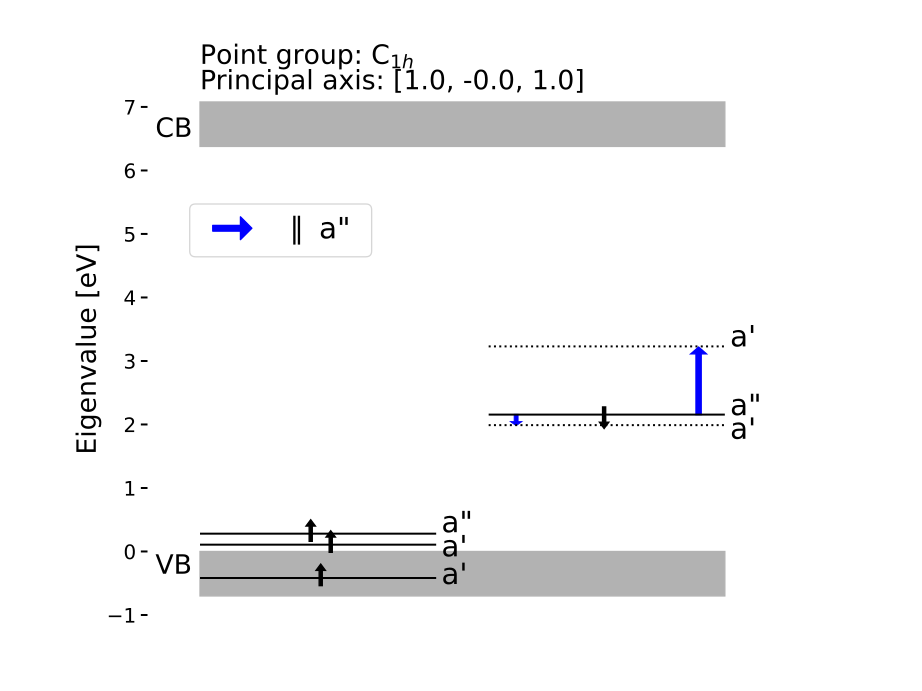} }}%
	\caption{Electronic structure of the $\mathrm{C_BV_B}$ defect. See Table~\ref{tab:cbvb_zpl} for ZPL using different k-points.}
	\label{fig:cbvb_transitions} 
\end{figure}

Figure~\ref{fig:cbvb_geo} shows the geometry of the $\mathrm{C_BV_B}$ defect and the spin density isosurface, the defect has $\mathrm{C_{1h}}$ symmetry in both ground and excited states. Figure~\ref{fig:cbvb_transitions} shows the electronic structure of the $\mathrm{C_BV_B}$ defect simulated at the $\Gamma$-point in the ground a) and excited b) state. The lowest energy excitation is between the occupied $a'$ state and the empty $a''$ state.
These simulations are using the regular 5.4.4 version of VASP, when the bug-fixed version of VASP \cite{PhysRevApplied.22.034056} is used in the excited state, the system relaxes back to the ground state.

\begin{table}[]
\caption{$\mathrm{C_BV_B}$ simulated at different k-points. 216 atom supercell. HSE with 33\% exact exchange.}
\begin{tabular} {c|c|l}
ZPL (eV) & k-point sampling & Ref. \\
\hline
 1.009 & $\Gamma$-point & \multirow{3}{1em}{This work} \\
 1.026 & 2x2x2 mesh using Monkhorst-Pack & \\
 0.952 & Special k-point (1/4, 1/4, 1/4) & \\
 \hline
 0.95 & Special k-point (1/4, 1/4, 1/4) & \cite{PhysRevB.108.L041102} \\
\end{tabular}
\label{tab:cbvb_zpl}
\end{table}

\bibliography{references}

\end{document}